\documentclass[english,a4paper]{article}

\usepackage{graphicx}
\usepackage{amsmath}
\usepackage{amssymb}
\usepackage{mathtools}
\usepackage{color}

\DeclareMathOperator{\diag}{diag}

\DeclareMathOperator{\K}{K}
\DeclareMathOperator{\F}{F}
\DeclareMathOperator{\am}{am}
\DeclareMathOperator{\arctanh}{arctanh}

\newcommand{\We}{\mathbf{E}}
\newcommand{\An}{\mathbf{J}}

\begin{document}
\title{Application of the Small Tip-Angle approximation in the Toggling Frame for the design of analytic robust pulses in Quantum Control}
\author{L. Van Damme\footnote{Department of Chemistry, Technical University of Munich, Lichtenbergstrasse 4, D-85747 Garching, Germany; leo.vandamme@gmx.fr}, D. Sugny\footnote{Laboratoire Interdisciplinaire Carnot de
Bourgogne (ICB), UMR 6303 CNRS-Universit\'e Bourgogne-Franche Comt\'e, 9 Av. A.
Savary, BP 47 870, F-21078 Dijon Cedex, France; dominique.sugny@u-bourgogne.fr}, S. J. Glaser\footnote{Department of Chemistry, Technical University of Munich, Lichtenbergstrasse 4, D-85747 Garching, Germany; Munich Center for Quantum Science and Technology (MCQST), Schellingstrasse 4, 80799 M\"unchen, Germany; glaser@tum.de}}

\maketitle

\begin{abstract}
We apply the Small Tip-Angle Approximation in the Toggling Frame in order to analytically design robust pulses against resonance offsets for state to state transfer in two-level quantum systems. We show that a broadband or a local robustness up to an arbitrary order can be achieved. We provide different control parameterizations to satisfy experimental constraints and limitations on the amplitude or energy of the pulse. A comparison with numerical optimal solutions is made.
\end{abstract}

\section{Introduction}
Manipulating quantum systems by means of time-dependent external controls has been a topic of increasing interest in the past decades. It has become a well-recognized field of research with applications ranging from molecular physics~\cite{Brumer03,RMP19,brif10} to Nuclear Magnetic Resonance (NMR)~\cite{glaser15,bernstein,nielsen2010,lapert_10,assemat_11,khaneja01} and nowadays quantum technologies~\cite{altafini,dong,glaser15}. In this context, progress has been made for the design of efficient pulses able to realize specific tasks. On the theoretical side, such advances extend from the discovery of elementary basic mechanisms of field-induced dynamics such as adiabatic protocols~\cite{RMPvitanov}, to shortcut to adiabaticity~\cite{RMPguery,deffner14,torrontegui12,martikyan20_2} and optimal control procedures~\cite{Pontryagin,bonnardbook,bryson,Dalessandro08,glaser15,reich12,Khaneja05,garon13}, which have made possible the control of systems of growing complexity. However, in order to be effective for experimental applications, such open-loop control methods require the accurate knowledge of system dynamics. This problem can be solved by considering pulses robust against variations of specific parameters of the system~\cite{li09}. The basic idea is generally to consider the simultaneous control of an ensemble of quantum systems which differ by the value of one or several parameters. A large amount of solutions have been proposed in the literature to date, with their own advantages and limitations~\cite{torosov_21} in terms of pulse duration and energy or efficiency of the control protocol. Among recent propositions for two-level quantum systems, we mention composite pulses~\cite{Li06,owrutsky12,genov_14,genov_20,jones13}, procedures based on shortcut controls~\cite{ruschhaupt12,Daems13,vandamme_17}, learning control~\cite{turinici19,chen14} and optimal control methods~\cite{kobzar04,kobzar05,kobzar12,lapert12_1,VanDamme17,glaser15,ruths11,ruths12,Ugurbil18}. As an illustrative example of this control issue, we consider in this study the control of two-level quantum systems with different resonance offsets, which can be viewed as a reference problem for robust protocols in quantum control.

In this context, the design of robust pulses is a non-trivial task due to the bilinearity of the controlled Schr\"odinger (or Bloch) equation. Ideally, the system dynamics and the corresponding control should be expressed in terms of simple functions with a minimum number of free parameters. This aspect is important to reveal the control mechanism or to apply quickly and efficiently the control protocols in a given experimental setup. However, the time evolution of the Schr\"odinger equation for two-level quantum systems can be analytically computed only for some simple controls, such as constant pulses with a constant phase. Most of the robust pulses have therefore been built on the basis of numerical optimizations of a large number of parameters, for which the integration of the dynamics is made by a numerical propagation. Analytical studies of the control of an ensemble of two-level quantum systems is much more difficult and requires in general some approximations to simplify the dynamics. For this purpose, Average Hamiltonian Theory (AHT)~\cite{Haeberlen76,Brinkmann16} uses a Magnus expansion to express the propagator. This expansion becomes extremely complicated above the second order, which limits the efficiency of this approach. The Small Tip-Angle approximation (STA)~\cite{Pauly89,bernstein} is another way to deal with the control problem and gives interesting results for state to state transfers. STA linearizes the Bloch equation which allows to compute explicitly its solutions. However, this method only works for transfers involving relatively small flip angles on the Bloch sphere, and not e.g. for an inversion process. The combination of STA and optimal control has been recently investigated in detail~\cite{Li17,martikyan20_2}. These two methods are much more efficient in the so-called \emph{Toggling Frame} (or Interaction Frame)~\cite{Haeberlen76,Brinkmann16}, hereafter denoted TF. The latter is a time dependent frame which follows the state of a resonant spin. Recently, Zeng et al. published a series of papers about the application of AHT in TF~\cite{Zeng18a,Zeng18b,Zeng19}. They found a way to improve the pulse robustness by making use of three-dimensional curves derived from the first order Magnus expansion. Their procedure allows to control the robustness of unitary gates by cancelling the effect of offset inhomogeneities in higher and higher orders of the Magnus expansion. Due to its complexity, the computation of high orders requires numerical techniques. Moreover, this method improves \emph{locally} the robustness (i.e. for small resonance offsets), but, as far as we know, broadband robust pulses for a large range of frequencies cannot be derived.

In this paper, we propose to revisit this approach by applying STA in TF. Many original results can be found for state to state transfers. This method has the decisive advantage of being simpler and more efficient than AHT, even if its generalization to unitary transformations seems more difficult. This paper focuses mainly on robust inversion pulses against offset inhomogeneities (or $B_0$ inhomogeneities), but we show that our approach can be generalized to any state to state transfer. The evaluation of the robustness is made through the distance of the final state to the target one as a function of the offset parameter relative to the resonance frequency. This description corresponds to the current experimental uncertainties that can be encountered in molecular physics, NMR or quantum technologies. We consider two different definitions of robustness, either global or local. In the first case,
the problem is to control an inhomogeneous ensemble of spins of different offsets, while in the second framework the system is expanded order by order with respect to the offset parameter.

The paper is organized as follows. In Sec.~\ref{sec1}, we introduce the model system and we express its dynamics in TF by using STA. We also describe the general methodology used to design robust control pulses. Section~\ref{secApplic} is mainly dedicated to the inversion process. We derive a series of analytical solutions both with broadband or local robustness properties. A generalization to any state to state transfer is presented in Sec.~\ref{SectArbitrarys2s}. A comparison with numerical optimal control protocols is presented in Sec.~\ref{numsec}. Conclusion and prospective views are given in Sec.~\ref{conclusion}. Technical details are reported in Appendices~\ref{AppEq23}, \ref{AppElliptic} and~\ref{AppPolynom}.
\section{Methodology\label{sec1}}
\subsection{The model system\label{secHI}}
We consider an inhomogeneous ensemble of uncoupled two-level quantum systems with different resonance offsets neglecting relaxation. In a given rotating frame, the dynamics of the Bloch vector describing the state of a system are given by:
\begin{equation}
\dot{\vec{M}}(\delta,t)=(H_0(t)+\delta H_1)\vec{M}(\delta,t)\label{eqBloch}
\end{equation}
with:
\begin{equation}
\begin{aligned}
& H_0(t)=\begin{pmatrix}
0 & 0 & -u_y(t) \\ 0 & 0 & u_x(t) \\ u_y(t) & -u_x(t) & 0
\end{pmatrix},\\
& H_1=\begin{pmatrix}
0 & 1 & 0\\
-1 & 0 & 0\\
0 & 0 & 0
\end{pmatrix},
\end{aligned}
\end{equation}
where $\delta$ is the resonance offset and $u_x(t)$ and $u_y(t)$ are the components of the control pulse. The control would be infinitely robust if $\vec{M}(\delta,T)$ reaches a given target state for all $\delta$. For an inversion pulse, the cost profile (i.e. the error of the transfer as a function of $\delta$) is measured by:
\begin{equation}
J_{\textsc{Bloch}}(\delta)=1+M_z(\delta,T),\label{eqCostBloch}
\end{equation}
which is zero if the inversion is perfectly realized for an offset $\delta$ (we have in this case $M_z(\delta,T)=-1$), and $2$ if the spin is not excited at all.
TF (or Interaction Frame) ~\cite{Brinkmann16,Levitt08,CohenBook} is defined by a propagator whose dynamics are governed by $H_0$ only. In other words, it corresponds to a rotation matrix  $R_0\in SO(3)$ which fulfills $\dot{R}_0=H_0 R_0$. $R_0$ is an orthogonal $3\times 3$ matrix that can be expressed as:
\begin{equation}
R_0(t)=\begin{pmatrix}
q_x(t)& q_y(t)& q_z(t)\\
p_x(t)& p_y(t)& p_z(t)\\
v_x(t)& v_y(t)& v_z(t)
\end{pmatrix},\label{eqRotMatrix}
\end{equation}
where we have introduced the vectors $\vec{q}=(q_x,q_y,q_z)^\intercal$, $\vec{p}=(p_x,p_y,p_z)^\intercal$, and $\vec{v}=(v_x,v_y,v_z)^\intercal$. These vectors corresponds to the three axes of the original frame expressed in the Toggling Frame. The vector $\vec{q}$ is the $x$-axis of the original frame expressed in TF, $\vec{p}$ is the $y$-axis and $\vec{v}$ is the $z$-axis. Since initially we have $R_0(0)=\mathbb{I}$ (the two frames are equal at the beginning of the process), we obtain $\vec{q}(0)=(1,0,0)^\intercal$, $\vec{p}(0)=(0,1,0)^\intercal$ and $\vec{v}(0)=(0,0,1)^\intercal$.
We denote as $\vec{L}$ the Bloch vector described in TF, i.e. such that its components $L_x$, $L_y$ and $L_z$ correspond to the projection of $\vec{M}$ onto the three axes of TF. The dynamics of $\vec{L}$ can be derived by applying the transformation $\vec{L}=R_0^\intercal\vec{M}$. Using Eq.~\eqref{eqBloch}, we obtain:
\begin{equation}
\dot{\vec{L}}(\delta,t)=\delta \tilde{H}_1(t) \vec{L}(\delta,t),\label{eqTogg}
\end{equation}
where $\tilde{H}_1=R_0^\intercal H_1 R_0$ is the so-called Interaction Hamiltonian. Since $R_0$ is an orthogonal matrix, we deduce that $\vec{q}\times\vec{p}=\vec{v}$ which allows to explicitly write $\tilde{H}_1$ as:
\begin{equation}
\tilde{H}_1(t)=\begin{pmatrix}
0 & v_z(t) & -v_y(t) \\ -v_z(t) & 0 & v_x(t) \\ v_y(t) & -v_x(t) & 0
\end{pmatrix}.\label{eqHI}
\end{equation}
Note that at the resonance, we have $\dot{\vec{L}}(\delta=0,t)=0$, i.e. the Bloch vector is a constant of the motion. This vector stays along the $z$- axis of TF during the dynamics. Since the dynamics of $R_0$ are controllable, the functions $v_x$, $v_y$ and $v_z$ can be chosen arbitrarily
and Eq.~\eqref{eqTogg} can be viewed as a new control problem, where $v_x$, $v_y$ and $v_z$ are the new control variables. Note that in the system~\eqref{eqTogg}, the Interaction Hamiltonian generates rotations about $\vec{v}$, hence $\vec{v}$ can be seen as a control pulse by analogy with the transverse control field in the original Bloch equation.
The fact that $R_0$ is orthogonal leads to the constraint:
\begin{equation}
v_x^2(t)+v_y^2(t)+v_z^2(t)=1.\label{eqvxyz}
\end{equation}
As we explained above, the vector $\vec{v}$ is also equal to the unit vector along the $z$-axis of the laboratory frame (see Fig.~\ref{Fig1}). It can be shown that the original pulse can be computed from $\vec{v}$ as~\cite{Zeng19}:
\begin{equation}
\Omega(t)=\Vert\dot{\vec{v}}(t)\Vert,\;\; \phi(t)=\int_0^t\frac{\big(\vec{v}(t')\times\dot{\vec{v}}(t')\big)\cdot\ddot{\vec{v}}(t')}{\Omega^2(t')}dt',\label{eqFieldFromv}
\end{equation}
where $\Omega=\sqrt{u_x^2+u_y^2}$ and $\phi=\arctan(u_y/u_x)$ are respectively the amplitude and the phase of the pulse. We stress that $\Omega$ is the norm of the derivative of $\vec{v}$, and not the derivative of the norm. Instead of optimizing directly $u_x(t)$ and $u_y(t)$, we can optimize $\vec{v}(t)$ in Eq.~\eqref{eqTogg} and~\eqref{eqHI} and deduce the pulse through Eq.~\eqref{eqFieldFromv}.

\subsection{Boundary constraints\label{secBoundConstr}}
The main difference between the new dynamical system and the original one is that the vector $\vec{v}(t)$ must satisfy boundary constraints at $t=0$ and $t=T$. This point is due to the fact that TF is not a static frame. It has to realize a certain transfer which depends on the target state of the control problem. TF is equal to the original frame at $t=0$, i.e. $R_0(0)=\mathbb{I}$ which, from Eq.~\eqref{eqRotMatrix}, leads to:
\begin{equation}
\vec{v}(0)=(0,0,1)^\intercal,
\end{equation}
while $\vec{v}(T)$ depends on the target state. For the design of an inversion pulse, TF has to be flipped, i.e. $R_0(T)=\diag(1,-1,-1)$, leading to:
\begin{equation}
\vec{v}(T)=(0,0,-1)^\intercal. \label{eqConstraintvT}
\end{equation}
A pulse is said to be robust against offset variations if for any offset $\delta$, the Bloch vector remains in a neighborhood of the $z$-axis of TF. As explained above, a resonant Bloch vector with $\delta=0$ stays exactly along the $z$-axis of TF during the control process, i.e. $\vec{L}(0,t)=(0,0,1)^\intercal$. Ideally, if, at the final time, we have $\vec{L}(\delta,T)=(0,0,1)^\intercal$ $\forall\delta$, then the associated pulse is infinitely robust. In other words, in TF, a robust control process steers the systems from the $z$- axis back to the $z$-axis. The final angle between the Bloch vector and the $z$-axis of TF is measured by $\arccos(L_z(\delta,T))$. The cost profile can thus be defined as follows:
\begin{equation}
J_{\textsc{TF}}(\delta)=1-L_z(\delta,T),\label{eqCost}
\end{equation}
which is zero for a perfect return to the $z$-axis ($L_z(\delta,T)=1$). It is worth noting that the cost does not depend explicitly on the target state, this latter being determined by the final constraint $\vec{v}(T)$. For an arbitrary target flip angle $\theta_T$ on the Bloch sphere, the constraint~\eqref{eqConstraintvT} can be generalized to $\vec{v}(T)=(\sin[\theta_T],0,\cos[\theta_T])^\intercal$ while the cost profile is still given by~\eqref{eqCost}. The azimuthal target angle does not appear in this constraint, but it can be set by adding an appropriate constant to the phase of the pulse.

In general, Eq.~\eqref{eqTogg} is not simpler to solve than the original Bloch equation and does not provide any advantage. However, some approximations can be made such as AHT which is much more accurate in TF~\cite{Brinkmann16,Haeberlen76}. Another solution is to use STA~\cite{Pauly89} which is particularly efficient for state to state transfers. We show in Sec.~\ref{secSTA} that it is a very natural approach in this control problem.
\subsection{Application of the Small Tip-Angle Approximation\label{secSTA}}
A large amount of works have used STA for the design of robust or selective pulses in NMR and Magnetic Resonance Imaging (MRI) due to its very good efficiency in state to state transfers. Among others, we mention the Small Tip Angle spokes~\cite{Grissom12}, $k_T$-point pulses~\cite{Cloos11,Gras18}, SPINS~\cite{Malik11} and Fast $k_z$- pulses~\cite{Saekho06}.
The unit vector $\vec{L}$ can be represented in polar coordinates as: $$\vec{L}=(\sin\alpha\cos\beta,\sin\alpha\sin\beta,\cos\alpha)^\intercal,$$ $\alpha$ being the flip angle and $\beta$ the azimuthal one. If $\alpha$ is small enough, we have $\cos\alpha\simeq 1$ and the vector $\vec{L}$ moves in a plane tangent to the sphere in $L_z=1$. We thus have $\vec{L}\simeq\vec{\ell}=(\ell_x,\ell_y,1)^\intercal$, which is however not of norm unity. If needed, the mapping between $\vec{\ell}$ and $\vec{L}$ is given by $\alpha=\sqrt{\ell_x^2+\ell_y^2}$ and $\beta=\arctan\left(\frac{\ell_y}{\ell_x}\right)$.

\begin{figure}[h!]
\centering
\includegraphics[scale=0.65]{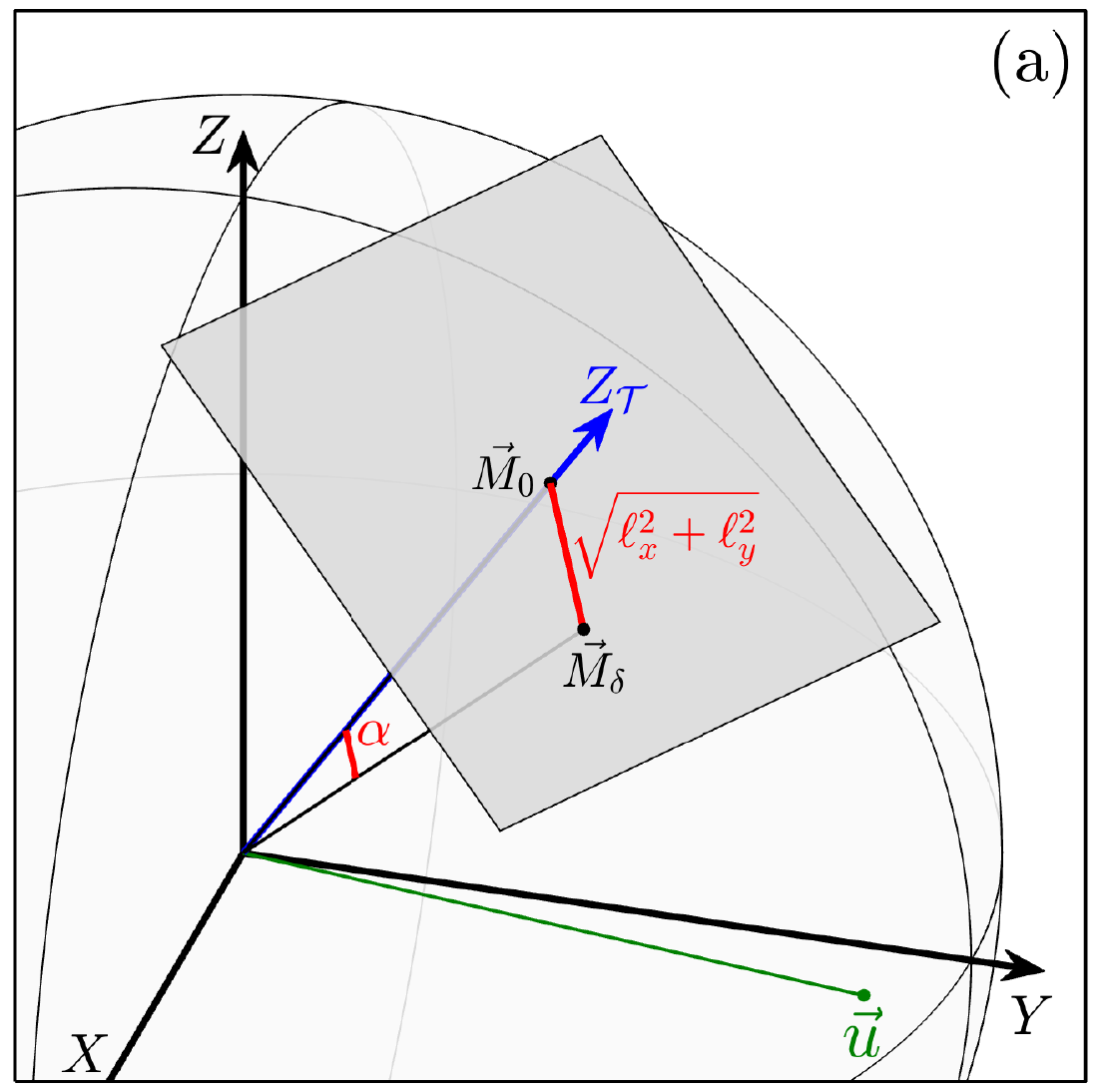}
\includegraphics[scale=0.65]{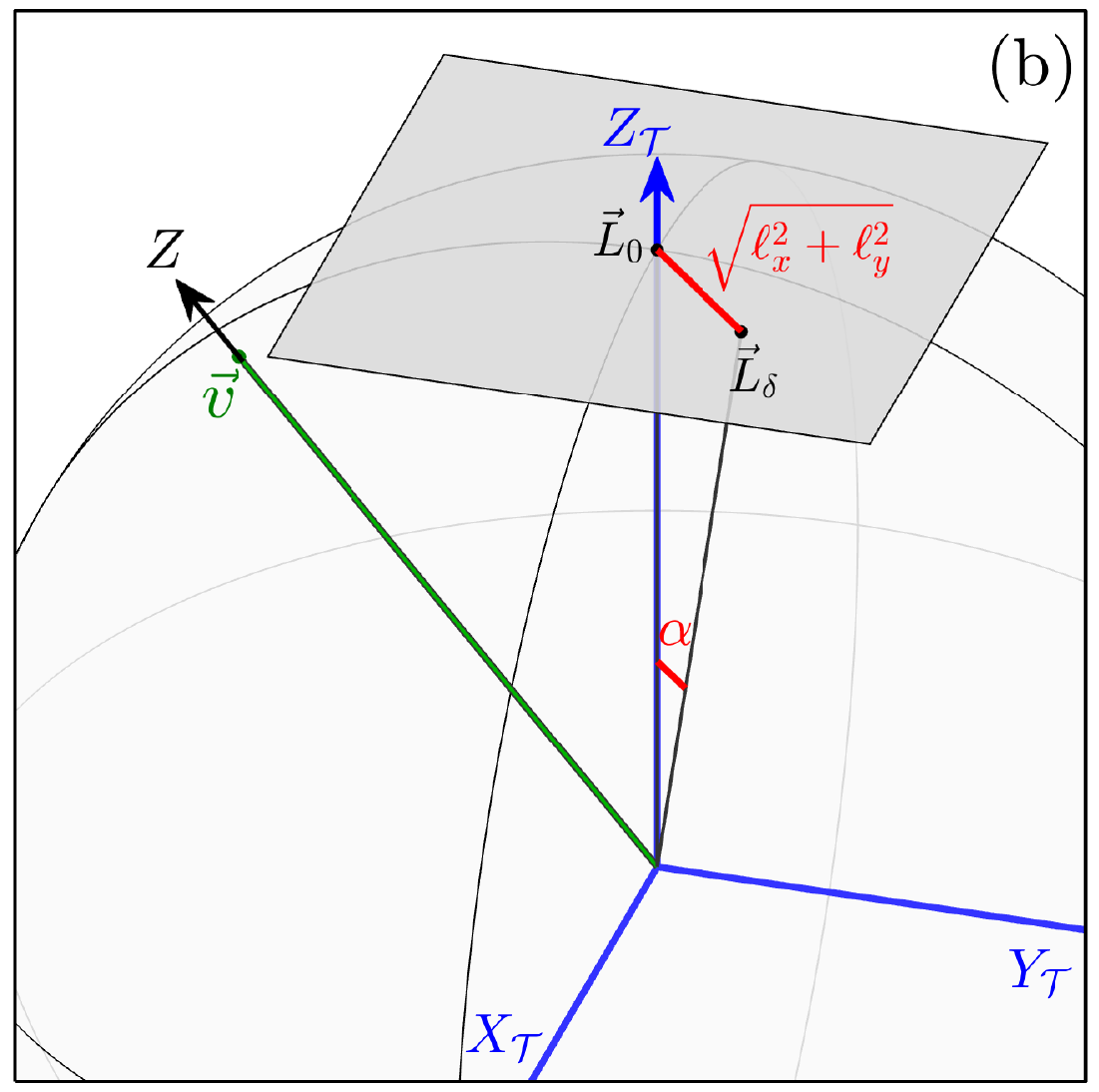}
\caption{\emph{Panel (a):} Representation of the $z$- axis of TF (${Z}_{\mathcal{T}}$ in blue) moving on the Bloch Sphere together with a resonant Bloch vector $\vec{M}_0\equiv\vec{M}(0,t)$ and a non-resonant one $\vec{M}_{\delta}\equiv\vec{M}(\delta,t)$. The vector $\vec{u}(t)$ is the transverse control $\vec{u}=(\Omega\cos\phi,\Omega\sin\phi,0)^\intercal$.
The resonant Bloch vector remains along ${Z}_{\mathcal{T}}$ during the whole process. The angle between ${Z}_{\mathcal{T}}$ and $\vec{M}(\delta,t)$ is denoted as $\alpha(t)$. If $\alpha$ is small enough during the dynamics, $\vec{M}(\delta,t)$ moves approximately in the gray plane and STA holds in TF. The angle $\alpha(t)$ is measured by $\sqrt{\ell_x^2(\delta,t)+\ell_y^2(\delta,t)}$ which is also the distance between ${Z}_{\mathcal{T}}$ and $\vec{M}(\delta,t)$ (red line). \emph{Panel (b)}: Representation of the $z$- axis of the laboratory frame ($Z$ in black) moving in TF. In this framework, the vector $\vec{v}(t)$ plays the role of a pulse (by analogy with $\vec{u}$ in the original frame) with a norm $\Vert\vec{v}\Vert=1$, and remains along $Z$ during the whole process. Here $\alpha(t)$ is the polar angle describing the dynamics of $\vec{L}(\delta,t)$.
The vectors $\vec{M}$ of panel (a) and $\vec{L}$ of panel (b) are the same but expressed in different frames. Dimensionless units are used.\label{Fig1}}
\end{figure}

STA can be applied for flip angles less than $30^{\circ}$~\cite{Pauly89}, but some studies use this approximation up to $105^{\circ}$, which still works surprisingly well~\cite{Pasteur,VanDamme21}. It is particularly relevant in TF. Since the resonant Bloch vector is static and stays along the $z$-axis of TF, we deduce that if $\delta$ is small enough, the corresponding vector stays in a neighborhood of the $z$-axis of TF, which means that the flip angle remains small. Since STA is valid even for relatively large flip angles, we expect that it can be applied over a relatively large range of offsets. Figure~\ref{Fig1} depicts the $z$- axis of TF, the tangent plane where the STA holds and the flip angle $\alpha$.

A decisive advantage of using STA in TF is that the solution of the Bloch equation can be expressed through a simple integral. The $z$-component of the Bloch vector $\vec{L}$ being equal to $1$, we deduce from Eq.~\eqref{eqTogg} that the dynamics of the $x$ and $y$-coordinates are given by:
\begin{equation}
\begin{aligned}
& \dot{\ell}_x(\delta,t)=\delta\big[v_z(t)\ell_y(\delta,t)-v_y(t)\big],\\
& \dot{\ell}_y(\delta,t)=\delta\big[v_x(t)-v_z(t)\ell_x(\delta,t)\big].
\end{aligned}
\end{equation}
We consider the complex component $\ell=\ell_x+i\ell_y$. We have:
\begin{equation}
\dot{\ell}(\delta,t)=-i\delta v_z(t)\ell(\delta,t)+i\delta[v_x(t)+iv_y(t)].
\label{eqdifflSTA}
\end{equation}
Introducing the functions:
\begin{equation}
\begin{cases}
k_x(t)=\int_0^tv_x(t')dt', \\
k_y(t)=\int_0^tv_y(t')dt', \\
k_z(t)=\int_0^tv_z(t')dt',
\end{cases}
\label{eqkxyz}
\end{equation}
one can check that the solution of~\eqref{eqdifflSTA} is of the form:
\begin{equation}
\ell(\delta,T)=i\delta e^{-i\delta k_z(T)} \int_0^T\big(\dot{k}_x(t)+i\dot{k}_y(t)\big)e^{i\delta k_z(t)}dt.
\label{eqSTAToggling}
\end{equation}
All the nonlinear properties of the Bloch equations are now contained in the functions $k_x$, $k_y$ and $k_z$.
In this system, the angle between the Bloch vectors and the $z$- axis of TF is measured by $\alpha=|\ell|$ with $|\ell|=\sqrt{\ell_x^2+\ell_y^2}$. In order to be consistent with Eq.~\eqref{eqCost}, we define the cost profile as:
\begin{equation}
J_{\textsc{STA}}(\delta)=1-\cos\big(|\ell(\delta,T)|\big).\label{eqCostSTA}
\end{equation}
This cost is nullified for $\vert\ell(\delta,T)\vert=0$. Note that it is also equal to $0$ when $\vert\ell(\delta,T)\vert$ is a multiple of $2\pi$. However, in this later case, the angle between the Bloch vectors and the $z$- axis of TF is too large to consider STA as a valid approximation.
Equation~\eqref{eqSTAToggling} is very similar to the master equation involved in $k$-space analysis in MRI~\cite{Pauly89}. The function $k_z$ plays here the role of a one dimensional $k$-space and $\dot{k}_x$ and $\dot{k}_y$ are analog to a RF-pulse. However, while in a standard application of STA, $k_x(t)$, $k_y(t)$ and $k_z(t)$ would be independent, here Eq.~\eqref{eqvxyz} involves that their derivatives satisfy $\dot{k}_x^2+\dot{k}_y^2+\dot{k}_z^2=1$ which might result in difficulties for solving analytically the integral~\eqref{eqSTAToggling}. A technique that allows to overcome this problem is presented in Sec.~\ref{secMethod}.
\subsection{Local robust control\label{secLocalControl}}
We also consider the robustness against local variations of the offset~\cite{Zeng19,Tycko84,Cummins03,Daems13,VanDamme17}. In this case, since $\delta\to 0$, the solution of the Bloch equation can be approximated by a Taylor series of the form: \[\vec{L}(\delta,t)=\vec{L}_0(t)+\delta\vec{L}_1(t)+\delta^2\vec{L}_2(t)+\cdots\]
Instead of minimizing a cost function over a certain range of offsets, all the vectors $\vec{L}_n$ can be canceled up to the order $n=N$ at the final time to ensure that the inhomogeneities do not disturb the system up to an error $\delta^N$. This results in a very good fidelity in a neighborhood of $\delta=0$, i.e. in a local robustness. The method is relevant for our problem since the precision of STA increases as $\delta$ becomes smaller. Its application is natural in this framework as it consists in truncating the generating series of the exponential of Eq.~\eqref{eqSTAToggling}, i.e.:
\begin{equation}
\small
\ell^{(N)}(\delta,T)=i\delta e^{-i\delta k_z(T)} \int_0^T\big(\dot{k}_x(t)+i\dot{k}_y(t)\big)\sum_{n=0}^{N-1}\frac{(i\delta k_z(t))^n}{n!}dt.
\end{equation}
A $N$-th order robust process is realized by nullifying both the real and imaginary part of this integral, i.e. by finding functions $k_x(t)$, $k_y(t)$ and $k_z(t)$ such that the $N$ following integrals cancel:
\begin{equation}
\small
C_n\equiv\int_0^T\big(\dot{k}_x(t)+i\dot{k}_y(t)\big)k_z^n(t)dt=0,\;\; n=\{0,\cdots,N-1\}.\label{eqCnLoc}
\end{equation}
Zeng et al. have analyzed the problem of local robustness in a series of papers~\cite{Zeng19,Zeng18a,Zeng18b} through 3D-curves that are in our case $\Gamma=(k_x,k_y,k_z)$. However, they focused on the geometric properties of these curves and not on the derivation of explicit analytic pulses. The resulting pulses can be of very high peak amplitude which may make them unrealistic experimentally. We use here a slightly different approach that allows us to find analytical expressions of the controls that are suitable for practical implementation in the case of the robust inversion problem.
\subsection{Analytical pulse design\label{secMethod}}
Our analytical study consists in finding some functions $k_x$, $k_y$ and $k_z$ that are suitable to integrate Eq.~\eqref{eqSTAToggling} in order to improve the cost profile~\eqref{eqCostSTA} or to cancel the integrals~\eqref{eqCnLoc}. The choice of these functions must be made very carefully.
Since the $k_i$'s are defined as integrals (see Eq.~\eqref{eqkxyz}), these functions are zero at $t=0$, i.e. $\vec{k}(0)=\vec{0}$. The boundary constraints on $\vec{v}$ described in Sec.~\ref{secBoundConstr}, which determine the transfer, imply that the time derivative of the functions $k_x$ $k_y$ and $k_z$ at $t=0$ fulfill $\dot{\vec{k}}(0)=(0,0,1)^\intercal$, and $\dot{\vec{k}}(T)=(0,0,-1)^\intercal$ for an inversion. An additional constraint discussed in Sec.~\ref{secSTA} is that $\Vert\dot{\vec{k}}\Vert=1$.

This latter restricts dramatically the choice of $\vec{k}(t)$. Finding a basis of functions satisfying $\Vert\dot{\vec{k}}\Vert=1$  is possible only in some simple cases.
This problem has been solved in Ref.~\cite{Zeng19,Zeng18a,Zeng18b} and we use here the same method. For any function $s$ increasing monotonously from $s(0)=s_0$ to $s(T)=s_T$, it is straightforward to show that the integral~\eqref{eqSTAToggling} can be rewritten as:
\begin{equation}
\ell(\delta,T)=i\delta e^{-i\delta k_z(s_T)}\int_{s_0}^{s_T}\left(\tfrac{d k_x}{d s}+i\tfrac{d k_y}{d s}\right)e^{i\delta k_z(s)}ds.
\label{eqSTAm2s}
\end{equation}
Therefore, one can choose some functions $k_x(s)$, $k_y(s)$ and $k_z(s)$ such that $\Vert d\vec{k}(s)/d s\Vert\neq 1$, while the function $s(t)$ is deduced from $\Vert d\vec{k}(t)/dt\Vert=1$. Indeed, we have:
\begin{equation}
\left\Vert \tfrac{d\vec{k}}{dt} \right\Vert=1
\Rightarrow \left\Vert \tfrac{d\vec{k}}{ds} \right\Vert \tfrac{ds}{dt}=1
\Rightarrow \left\Vert \tfrac{d\vec{k}}{ds} \right\Vert ds=dt.\label{eqdssurdt}
\end{equation}
Integrating from $s_0$ to $s$, we obtain:
\begin{equation}
\int_{s_0}^{s}\big\Vert \tfrac{d\vec{k}(s')}{d s'}\big\Vert ds'=t,\label{eqs2t}
\end{equation}
while the duration $T$ of the pulse is given by integrating until $s=s_T$.
The function $s(t)$ is thus obtained by inverting this integral, which can be done numerically if necessary. The boundary constraints are also slightly relaxed. For an inversion, it is now sufficient to select functions such that:
\begin{equation}
\begin{aligned}
& \tfrac{d k_x}{d s}=0,\; \tfrac{d k_y}{d s}=0,\;\tfrac{d k_z}{d s}> 0\text{ at }s=s_0,\\
& \tfrac{d k_x}{d s}=0,\; \tfrac{d k_y}{d s}=0,\;\tfrac{d k_z}{d s}< 0\text{ at }s=s_T
\end{aligned}\label{eqBoundConstr2sInv}
\end{equation}
Finally, we show in appendix~\ref{AppEq23} that the control pulses given by~\eqref{eqFieldFromv} can be expressed using the formula:
\begin{equation}
\begin{aligned}
&\Omega(s(t))=\left\Vert \tfrac{d\vec{k}}{d s}\times\tfrac{d^2\vec{k}}{d s^2}\right\Vert\left\Vert \tfrac{d\vec{k}}{d s}\right\Vert^{-3},\\
&\phi(s(t))=\int_{s_0}^{s(t)}\tfrac{\left(\frac{d\vec{k}}{d s}\times\frac{d^2\vec{k}}{d s^2}\right)\cdot\frac{d^3\vec{k}}{d s^3}}{\left\Vert \frac{d\vec{k}}{d s}\times\frac{d^2\vec{k}}{d s^2}\right\Vert^2} \left\Vert \tfrac{d\vec{k}}{d s}\right\Vert ds.
\end{aligned}\label{eqFieldFromk2s}
\end{equation}
The advantage of this expression is that it can be derived without knowing explicitly $s(t)$. The problem is thus simplified to the search of some functions $k_x(s)$, $k_y(s)$ and $k_z(s)$ under the constraints~\eqref{eqBoundConstr2sInv}. The derivative $d\vec{k}/ds$ does not need to belong to $S^2$. The pulse is given by~\eqref{eqFieldFromk2s} and the function $s(t)$ by~\eqref{eqs2t}.
\section{Application for inversion pulses\label{secApplic}}
\subsection{Broadband pulses}
\begin{figure*}
\centering
\includegraphics[scale=0.62]{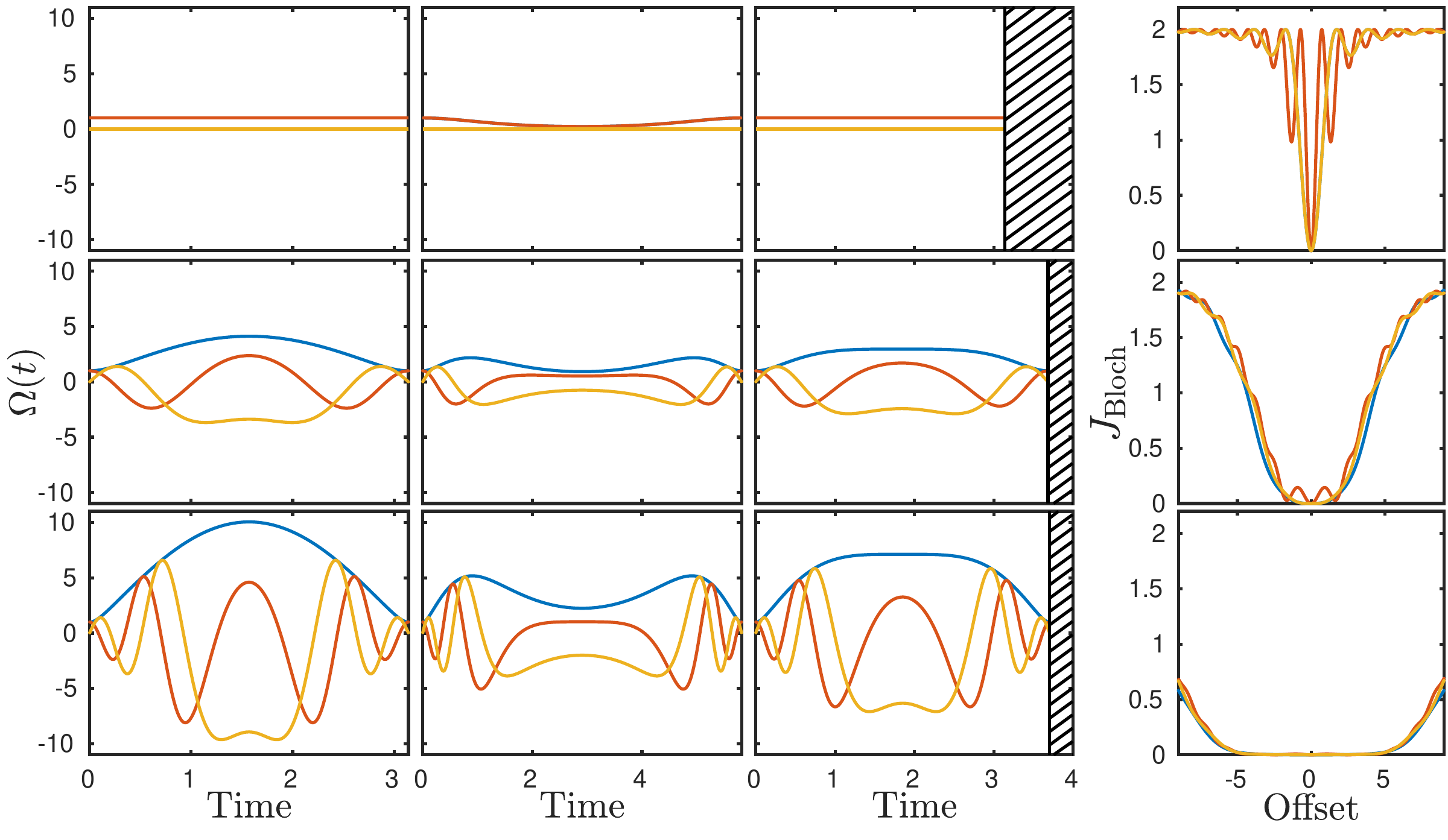}
\caption{\emph{First column:} Anger-Weber (or Jacobi with $m=0$) pulse amplitude  $\Omega(t)$ (blue line), $x$-component $\Omega\cos\phi$ (red line) and $y$-component $\Omega\sin\phi$ (yellow line) computed with Eq.~\eqref{PulseAnalyt} for various values of $\nu$. \emph{Second column:} Jacobi pulse amplitude (Eq.~\eqref{PulseJacobi}) and components computed for $m=0.95$. \emph{Third cloumn:}
Jacobi pulse amplitude and components computed for $m=\nu^2/(2\nu^2+1)$. From top to bottom, the values of $\nu$ are $\nu=0$, $\nu=4$ and $\nu=10$, respectively. \emph{Last column:}
 Associated cost profile computed from Eq.~\eqref{eqCost} for the Anger-Weber pulse (blue), the Jacobi pulse with $m=0.95$ (red) and the Jacobi pulse with $m=\nu^2/(2\nu^2+1)$ (yellow). Dimensionless units are used.\label{Fig2}}
\end{figure*}

Roughly speaking, Eq.~\eqref{eqSTAm2s} suggests that if $d k_x/d s$ and $d k_y/d s$ are two fast oscillating functions, the integral~\eqref{eqSTAm2s} is close to zero as well as the cost profile~\eqref{eqCostSTA}. The main idea of this section is to define a parameter, hereafter referred to as $\nu$, which determines the oscillating frequency of these functions. We expect that the robustness of the pulse increases with $\nu$.
In the following, we propose various parameterizations of $\vec{k}(s)$ leading to different types of pulses. For each of them, the robustness is verified using a numerical integration of the original Bloch equation~\eqref{eqBloch} and a computation of the cost profile~\eqref{eqCostBloch}.
In all the following results, the bounds of $s(t)$ are given by:
\begin{equation}
s_0=0 \longrightarrow s_T=\pi.
\end{equation}
An infinity of solutions could be derived. Only a few are described in this paper.
We stress that the technical computations are not straightforward. The use of a symbolic computation software such as \textsc{mathematica}~\cite{Mathematica}, \textsc{xmaxima}~\cite{maxima} or \textsc{maple}~\cite{maple} is particularly helpful.
\paragraph*{Anger-Weber solution.}
In this paragraph, we present all the steps of the method for the design of a simple analytic inversion pulse. Let us consider some functions $k_x(s)$, $k_y(s)$ and $k_z(s)$ satisfying:
\begin{equation}
\begin{aligned}
&\frac{d k_x}{d s}=\sin (s)\cos(\nu s)\\
&\frac{d k_y}{d s}=\sin (s)\sin(\nu s)\\
&\frac{d k_z}{d s}=\cos (s)\\
&k_z(s)=\sin (s)
\end{aligned}
\end{equation}
where $\nu$ is an arbitrary parameter introduced above that sets the oscillating frequency of the function. We can show that the complex transverse component given by Eq.~\eqref{eqSTAm2s} reads:
\begin{equation}
\small
\begin{aligned}
&\ell(\delta,T)=i\delta\int_{0}^\pi\sin(s)e^{i\nu s}e^{i\delta\sin(s)}ds\\
&=\frac{\delta}{2}\int_0^{\pi}\left(e^{i[(\nu+1)s+\delta\sin(s)]}-e^{i[(\nu-1)s+\delta\sin(s)]}\right) ds\\
&=-\left.\left.\frac{\delta\pi e^{i\nu\pi}}{2}\right[\An_{\nu+1}(\delta)-\An_{\nu-1}(\delta)-i\big(\We_{\nu+1}(\delta)-\We_{\nu-1}(\delta)\big)\right],
\end{aligned}
\end{equation}
where $\An$ and $\We$ are the Anger and Weber functions, respectively~\cite{Abramowitz}.
An interesting characteristic of these functions is that when $\nu$ increases, both $\We_\nu(\delta)$ and $\An_\nu(\delta)$ get closer to $0$ over a wider range of $\delta$. Since the cost profile is given by $1-\cos(|\ell(\delta,T)|)$ in STA, the parameter $\nu$ can be used to improve the robustness, as expected (see Fig.~\ref{Fig2}).

\begin{figure}[h!]
\centering
\includegraphics[scale=0.8]{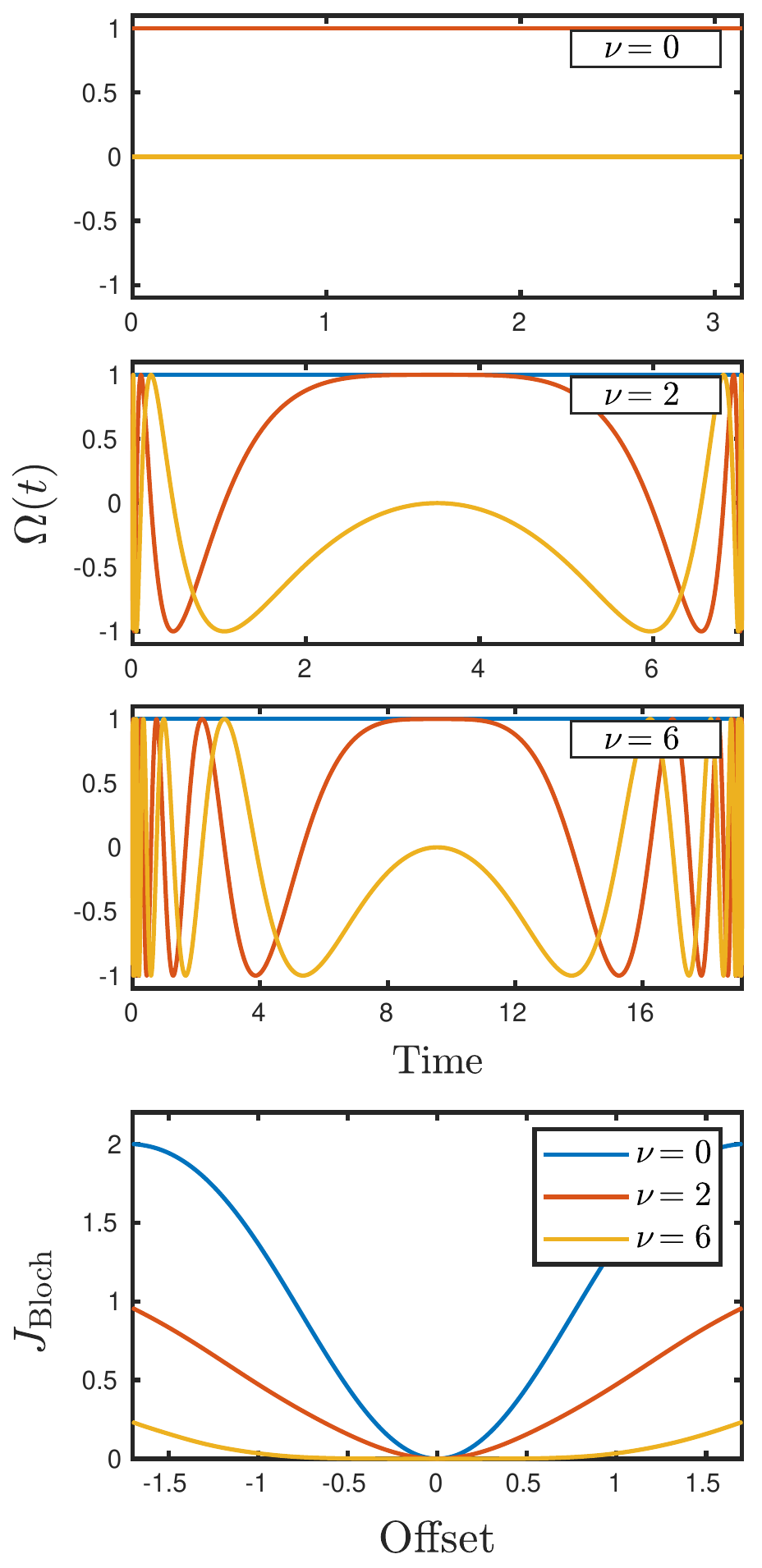}
\caption{\emph{Three upper pannels:} Amplitude-fixed pulse given by Eq.~\eqref{PulseSquare} for various values of $\nu$. The blue line represents the amplitude $\Omega(t)=1$, the red line $\Omega\cos\phi$ and the yellow line $\Omega\sin\phi$. \emph{Lower pannel:} Cost profile associated to each pulse. Dimensionless units are used.\label{Fig3}}
\end{figure}

Let us compute the corresponding pulse. The function $s(t)$ is given by Eq.~\eqref{eqs2t}, which leads to:
\begin{equation}
s(t)=t,
\end{equation}
and the total duration is $T=\pi$. The amplitude $\Omega$ and the phase $\phi$ are derived by using Eq.~\eqref{eqFieldFromk2s}. We used the free software \textsc{xmaxima} to simplify the latter formula. We obtain:
\begin{equation}
\begin{cases}
\Omega(t)=\sqrt{1+\nu^2\sin^2t},\\
\phi(t)=\nu\sin t+\arctan\left(\nu\sin t\right).
\end{cases}\label{PulseAnalyt}
\end{equation}
Note that in the limit $\nu=0$, the pulse is a simple square $\pi-$pulse of amplitude $1$. When $\nu$ tends to $\infty$, this pulse becomes infinitely robust.  The left panels of figure~\ref{Fig2} displays the pulse and its efficiency for different values of $\nu$.
\paragraph*{Jacobi pulse.}
The preceding solution has the advantage of being simple, but it is also interesting to derive solutions depending on more parameters in order to shape the pulse. The following result is slightly more general. The functions $k_x$, $k_y$ and $k_z$ are chosen such that:
\begin{equation}
\begin{aligned}
&\frac{d k_x}{d s}=\tfrac{\sin s\cos(\nu s)}{\sqrt{1-m\sin^2s}},\\
&\frac{d k_y}{d s}=\tfrac{\sin s\sin(\nu s)}{\sqrt{1-m\sin^2s}},\\
&\frac{d k_z}{d s}=\tfrac{\cos s}{\sqrt{1-m\sin^2s}},\\
&k_z(s)=\frac{\arcsin[\sqrt{m}\sin(s)]}{\sqrt{m}},
\end{aligned}
\end{equation}
where $m$ is an arbitrary modulus with $m\in[0,1]$.
Here, the transverse components~\eqref{eqSTAm2s} cannot be expressed in terms of simple functions. However, the pulse can be explicitly derived and its performances are computed numerically by integrating Eq.~\eqref{eqBloch}. It can be shown that the function $s(t)$ and the pulse are given by:
\begin{equation}
\begin{cases}
s(t)=\am(t,m), \\
\tfrac{ds}{dt}=\sqrt{1-m\sin^2[s(t)]},\\
\Omega(t)=\sqrt{1-m\sin^2 [s(t)]}\sqrt{1+\nu^2\sin^2 [s(t)]},\\
\phi(t)=\nu\sin [s(t)]+\arctan\left(\nu\sin [s(t)]\right),\\
T=2\K(m),
\end{cases}\label{PulseJacobi}
\end{equation}
where $\K(m)$ is a complete elliptic integral of the first kind, and $\am(t,m)$ is the Jacobi Amplitude~\cite{Abramowitz}. In this solution, the parameter $\nu$ still allows to improve the robustness profile, while the modulus $m$ is used to shape the pulse amplitude. Note that the case $m=0$ leads to the Anger-Weber solution computed in the previous paragraph.
Figure~\ref{Fig2} displays some of these pulses together with their performance. As can be seen, the parameter $m$ allows to reduce the maximum amplitude while increasing the time. The particular choice $m=\nu^2/(2\nu^2+1)$ involves that $\Omega(t)$ is very flat about $t=T/2$. The reason is that this choice cancels the second derivative of $\Omega(t)$ at $t=T/2$.
The shape of the cost profile is also affected by $m$. A deeper study would be necessary to choose a suitable $m$ for a given offset range and specific experimental constraints.
\paragraph*{Generalized Jacobi pulse.}
Many more parameters can be used to shape the pulse. We consider the functions $k_x$, $k_y$ and $k_z$ defined as:
\begin{equation}
\begin{aligned}
&\frac{d k_x}{d s}=\tfrac{\sin s\cos(\nu s)}{P_N(s)}\\
&\frac{d k_y}{d s}=\tfrac{\sin s\sin(\nu s)}{P_N(s)}\\
&\frac{d k_z}{d s}=\tfrac{\cos s}{P_N(s)},
\end{aligned}
\end{equation}
where the function $P_N$ is given by:
\begin{equation}
\footnotesize
P_N(s)=\sqrt{1-m_1\sin^2s}\sqrt{1-m_2\sin^2s}\cdots\sqrt{1-m_N\sin^2s},
\end{equation}
and $m_1,\cdots, m_N$ are arbitrary moduli belonging to $[0,1]$. The pulse can be expressed as:
\begin{equation}
\begin{cases}
s(t)=\am_N(t,m_1,\cdots, m_N), \\
\tfrac{ds}{dt}=P_N[s(t)],\\
\Omega(t)=P_N[s(t)]\sqrt{1+\nu^2\sin^2 [s(t)]},\\
\phi(t)=\nu\sin [s(t)]+\arctan\left(\nu\sin [s(t)]\right),\\
T=2\K_N(m),
\end{cases}
\end{equation}
where $\am_N(t,m_1,\cdots m_N)$ is the generalized Jacobi amplitude and $\K_N$ the generalized elliptic integral of the first kind (see Ref.~\cite{Pawellek11} for a complete description of these functions in the case $N=2$ and Appendix~\ref{AppElliptic} for the necessary properties). Again, increasing $\nu$ improves the robustness, while the extra parameters $m_i$ can be used to shape the pulse.
\paragraph*{Amplitude-fixed pulse.}
From a practical point of view, it is often necessary to have a pulse with a constant amplitude set by the experimental setup. This constraint can be satisfied by choosing the following parametrization of the functions $k_i$:
\begin{equation}
\begin{aligned}
&\frac{d k_x}{d s}=\sqrt{1+\nu^2}\sin s\cos\left[\nu\ln\left(\tan\left(\tfrac{s}{2}\right)\right)\right]\\
&\frac{d k_y}{d s}=\sqrt{1+\nu^2}\sin s\sin\left[\nu\ln\left(\tan\left(\tfrac{s}{2}\right)\right)\right]\\
&\frac{d k_z}{d s}=\sqrt{1+\nu^2}\cos s.
\end{aligned}
\end{equation}
Indeed, applying equations~\eqref{eqFieldFromk2s} lead to a pulse of the form:
\begin{equation}
\begin{cases}
s(t)=t/\sqrt{1+\nu^2}, \\
\Omega(t)=1,\\
\phi(t)=\nu\ln[\sin[s(t)]],\\
T=\pi\sqrt{1+\nu^2}.
\end{cases}\label{PulseSquare}
\end{equation}
The parameter $\nu$ improves the robustness by changing only the duration and the phase of the pulse, while keeping the amplitude equal to $1$.  Figure~\ref{Fig3} depicts some pulses and their performance. The case $\nu=0$ is a standard square $\pi$- pulse which can be used as a reference for comparing the efficiency of the different pulses.
Note that the phase is a fast oscillating function at the beginning and the end of the pulse. If necessary, these parts can be truncated while keeping a good robustness.

Many more solutions, maybe simpler or more efficient, could be derived easily using this method. As long as the functions $k_x$, $k_y$ and $k_z$ satisfy the boundary constraints, the inversion is realized at the resonance, i.e. for $\delta=0$. Since many functions satisfy the boundary constraints, we can construct an infinity of pulses with robust properties.
\subsection{Local robustness and orthogonal polynomial solutions\label{secChebyshev}}
The pulses presented in the previous section have been derived by guessing that if $dk_x/ds$ and $dk_y/ds$ oscillate with a high frequency, the integral~\eqref{eqSTAm2s} is small over a large range of offset, leading to a good robustness profile. Let us now consider the problem of local robustness that has been introduced in Sec.~\ref{secLocalControl}. We show that this problem can be solved by using \emph{Orthogonal Polynomials} in the interval $[-1,1]$ which have the following property:
\begin{equation}
\int_{-1}^1 \nu(x)p_n(x)p_m(x)dx=0\text{ if } n\neq m,\label{eqOrtPol}
\end{equation}
where $\nu(x)$ is called the weight function and $p_n$ is a polynomial of degree $n$. While many families of orthogonal polynomials could be used,
we prefer to focus on one solution based on Chebyshev polynomials of the first and second kinds. The function $s$ fulfills:
\begin{equation}
s_0=-1\rightarrow s_T=1,
\end{equation}
and we consider functions $k_x$, $k_y$ and $k_z$ that satisfy:
\begin{equation}
\begin{aligned}
&\frac{d k_x}{d s}=(1-s^2)\sqrt{1-s^2}U_{2n}(s),\\
&\frac{d k_y}{d s}=(1-s^2)T_{2n+1}(s),\\
&\frac{d k_z}{d s}=-2s,\\
&k_z(s)=1-s^2,
\end{aligned}\label{eqdkCheb}
\end{equation}
where $T_n$ and $U_n$ are Chebyshev polynomials of the first and second kinds, respectively~\cite{Abramowitz}.
Note that the boundary constraints~\eqref{eqBoundConstr2sInv} are satisfied for an inversion process. The transverse term given by Eq.~\eqref{eqSTAm2s} becomes:
\begin{equation}
\small
\begin{split}
&\ell(\delta,T)=\\
&i\delta\int_{-1}^1(1-s^2)\left(\sqrt{1-s^2}U_{2n}(s)+iT_{2n+1}(s)\right)e^{i\delta(1-s^2)}ds.
\end{split}
\end{equation}
Considering $\delta$ as a small perturbation, the exponential function can be truncated up to an arbitrary order $N$. We obtain:
\begin{equation}
\small
\begin{aligned}
& \ell^{(N)}(\delta,T)=\\
&\sum_{k=0}^{N-1}\frac{(i\delta)^{k+1}}{k!} \int_{-1}^1\left(\sqrt{1-s^2}U_{2n}(s)+iT_{2n+1}(s)\right)(1-s^2)^{k+1}ds.
\end{aligned}
\end{equation}
The problem is then to cancel $N$ integrals that are given by:
\begin{equation}
C_k=\int_{-1}^{1}\left(\sqrt{1-s^2}U_n(s)+iT_{n+1}(s)\right)(1-s^2)^{k}ds,\label{eqPolynomInt}
\end{equation}
with $k=\{1,\cdots,N\}$. Note that each term $(1-s^{2})^k$ is a symmetric function on $[-1,1]$. Moreover, the polynomial $T_{2n+1}(s)$ is antisymmetric, which involves that the imaginary part of $C_k$ cancels for all $k$. Since $(1-s^{2})^k$ is a polynomial of order $2k$ and that $U_{2k}$ is also a polynomial of order $2k$, each term $(1-s^{2})^k$ can be expressed as a linear combination of $\{U_{2k},U_{2k-1},\cdots,U_0\}$. In other words, each integral~\eqref{eqPolynomInt} can be written as:
\begin{equation}
C_k=\sum_{\ell=1}^{2k}a_\ell\int_{-1}^{1}\sqrt{1-s^2}U_{2n}(s)U_{\ell}(s)ds,
\end{equation}
for $k=\{1,\cdots,N\}$ and where the $a_\ell$'s are some coefficients that can be derived using a Chebyshev expansion, which is not necessary here. The weight function of the Chebyshev Polynomials $U_k(s)$ is given by $\sqrt{1-s^2}$. Therefore, using the orthogonality property~\eqref{eqOrtPol} and choosing $n=N+1$, all the terms of the sum cancel for every $k$, i.e.:
\begin{equation}
C_k=0\quad \forall k\in\{1,\cdots,N\},
\end{equation}
and the problem is solved.

\begin{figure}[h!]
\centering
\includegraphics[scale=0.6]{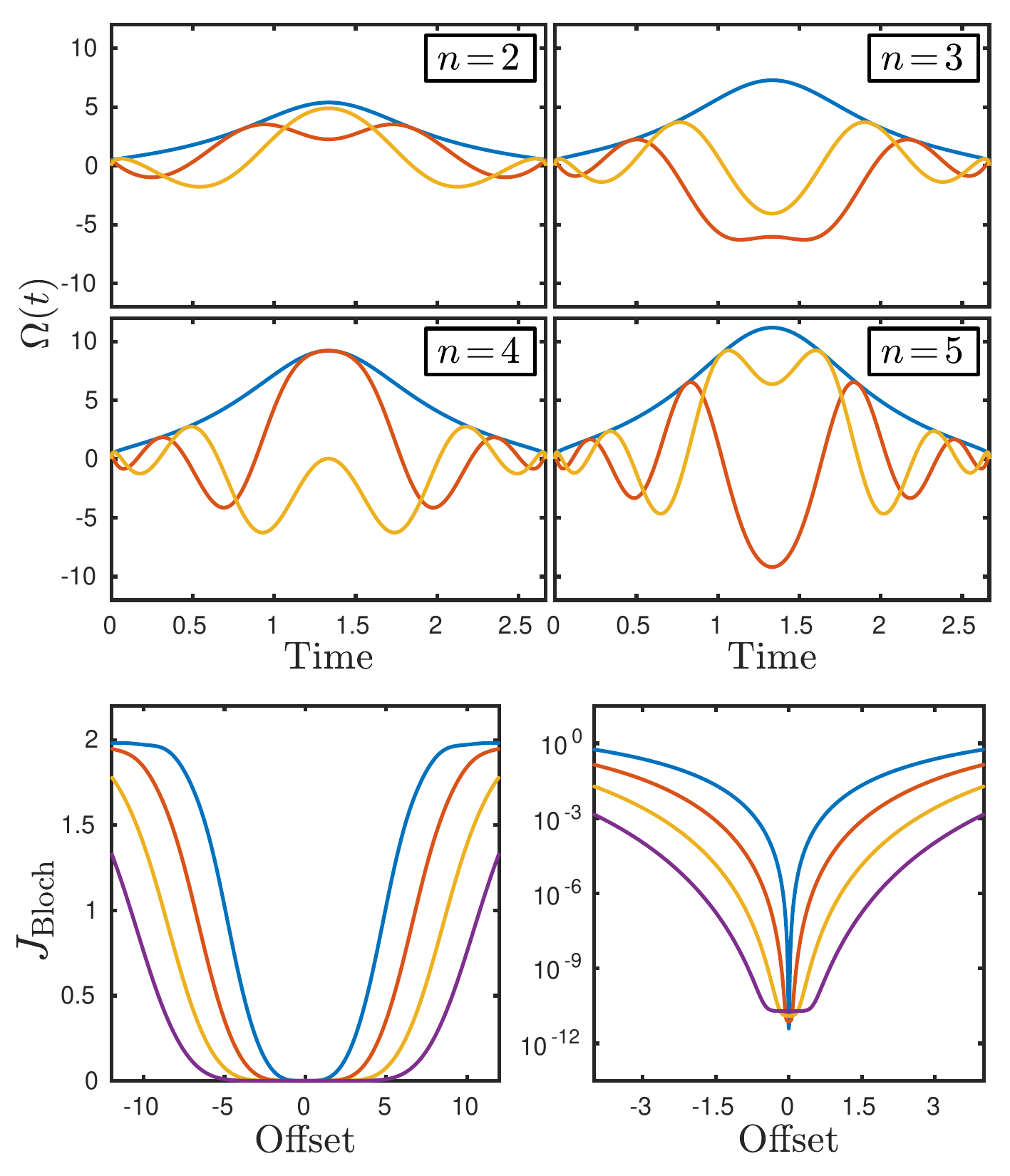}
\caption{\emph{Upper panels:} Pulse amplitude $\Omega(t)$ (blue line), $x$- component $\Omega\cos\phi$ (red line) and $y$- component $\Omega\sin\phi$ (yellow line) computed with Eq.~\eqref{eqChebyshevPulse} for $n=2$, $3$, $4$ and $5$. \emph{Lower panels:} Cost computed with Eq.~\eqref{eqCost} associated to $n=2$ (blue), $n=3$ (red), $n=4$ (yellow) and $n=5$ (purple). The right panel represents the profile in a logarithmic scale. Dimensionless units are used.
\label{fig4}}
\end{figure}

The final step consists in computing the associated pulse. The details of the computation can be found in Appendix~\ref{AppPolynom}. We obtain:
\begin{equation}
\small
\begin{cases}
s(t)=\left[\sqrt{1+\tfrac{(4-3t)^2}{4}}-\tfrac{4-3t}{2}\right]^{\tfrac{1}{3}}-\left[\sqrt{1+\tfrac{(4-3t)^2}{4}}+\tfrac{4-3t}{2}\right]^{\tfrac{1}{3}},\\
\Omega(t)=\tfrac{\sqrt{4+(2n+1)^2[1-s^2(t)]}}{[1+s^2(t)]^2},\\
\begin{split}
\phi(t)=&\sqrt{2}(2n+1)\arctanh\left(\sqrt{\tfrac{{1-s^2(t)}}{2}}\right)\\
&+\arctan\left(\tfrac{(2n+1)\sqrt{1-s^2(t)}}{2}\right),
\end{split}\\
T=8/3.
\end{cases}
\label{eqChebyshevPulse}
\end{equation}
Choosing $n=2$ for this pulse cancels the first order perturbation term, i.e. the integral $C_1$. For $n=3$, the integrals $C_1$ and $C_2$ are nullified, while in the case $n=4$, $C_1$, $C_2$ and $C_3$ are zero, and so on. Generally, choosing $n=N+1$ cancels the integrals $C_1$ to $C_N$.
The robustness is then locally improved up to an arbitrary order. Figure~\ref{fig4} shows the pulse for different values of $n$ and the associated cost computed by propagating numerically the original Bloch equation~\eqref{eqBloch}. A linear and a logarithmic scale is used for the cost profile in order to emphasize the very high precision of the transfer close to $\delta=0$.
\subsection{Arbitrary flip angle excitation pulses\label{SectArbitrarys2s}}
In this section, we generalize the method of local robustness to any flip angle excitation transfer. As explained in Sec.~\ref{sec1}, the transfer is fixed by the final constraint $\vec{v}(T)$. For a target flip angle $\theta_T$, the constraint is $\vec{v}(T)=(\sin[\theta_T],0,\cos[\theta_T])^\intercal$. With the change of variable $t\rightarrow s(t)$, the constraints~\eqref{eqBoundConstr2sInv} become:
\begin{equation}
\begin{aligned}
& \tfrac{d k_x}{d s}=0,\; \tfrac{d k_y}{d s}=0,\;\tfrac{d k_z}{d s}> 0\text{ at }s=s_0,\\
& \tfrac{d k_x}{d s}=A\sin\theta_T,\; \tfrac{d k_y}{d s}=0,\;\tfrac{d k_z}{d s}=A\cos\theta_T\text{ at }s=s_T,
\end{aligned}
\end{equation}
where $A$ is an arbitrary multiplicative constant coming from the fact that $\Vert d\vec{k}/ds\Vert$ does not need to be equal to $1$, unlike $\Vert\vec{v}\Vert$. We define the bounds of the function $s$ as:
\[s_0=-1\rightarrow s_T=1.\]
For the inversion process, the pulse symmetry allows us to derive analytically the pulse. For other state to state transfers, this trick cannot be used and both $d k_x/d s$ and $d k_y/d s$ have to be chosen such that each integral~\eqref{eqCnLoc} cancels due to the orthogonal properties of the polynomials (in Sec.~\ref{secChebyshev}, $d k_y/d s$ is antisymmetric which ensures that the imaginary part of the integrals cancels). Additional difficulties then appear in computing the pulse, as shown below.

\begin{figure}[h!]
\centering
\includegraphics[scale=0.8]{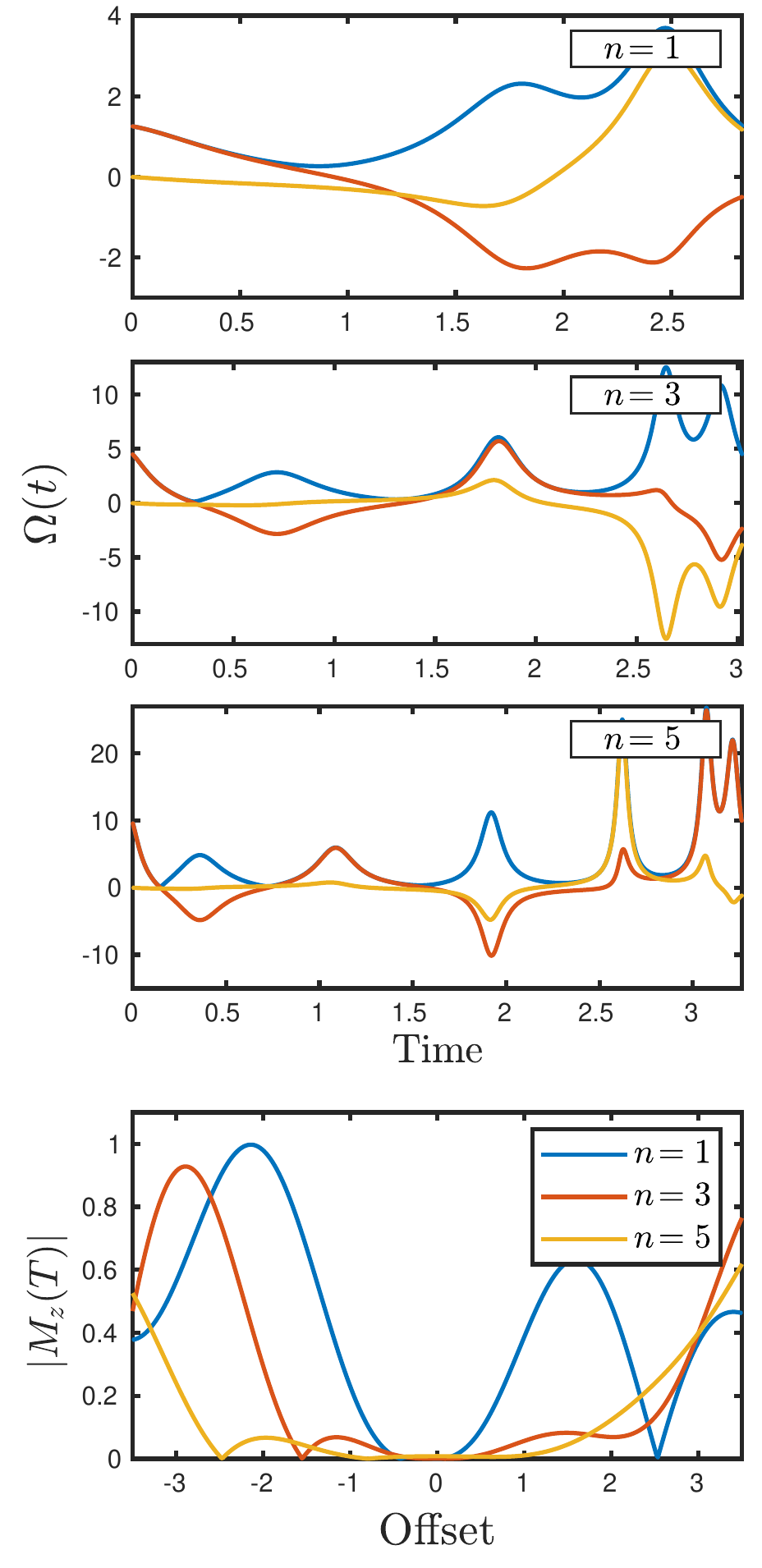}
\caption{Excitation pulse ($\theta_T=90^{\circ}$) robust up the first ($n=1$), second ($n=3$) and third ($n=5$) orders and associated excitation cost profile. Dimensionless units are used.\label{Fig5}}
\end{figure}

Let us choose the functions $k_x$, $k_y$ and $k_z$ such that:
\begin{equation}
\begin{aligned}
&\frac{d k_x}{d s}=\tfrac{(1+s)\sin\theta_T}{2}\left(P_{n}^{(0,1)}(s)+P_{n+1}^{(0,1)}(s)\right),\\
&\frac{d k_y}{d s}=\tfrac{(1+s)(n+1)}{2}\left(P_{n}^{(0,1)}(s)-P_{n+1}^{(0,1)}(s)\right)\\
&\frac{d k_z}{d s}=2\left(\cos^2\left(\tfrac{\theta_T}{2}\right)-s\sin^2\left(\tfrac{\theta_T}{2}\right)\right)\\
&k_z(s)=(1+s)^2\cos^2\left(\tfrac{\theta_T}{2}\right)+1-s^2,
\end{aligned}
\end{equation}
where $P_n^{(a,b)}$ is a Jacobi polynomial~\cite{Abramowitz}. The weight function of this polynomial being $(1-s)^a(1+s)^b$, we can show that a $N$-th order robust control can be found by choosing:
\begin{equation}
n= 2N-1.
\end{equation}
The function $s(t)$ cannot be explicitly found in this case because it is given by the inverse of the following integral:
\begin{equation}
\begin{aligned}
{dt}=&\sqrt{\left(\tfrac{d k_x}{d s}\right)^2+\left(\tfrac{d k_y}{d s}\right)^2+\left(\tfrac{d k_z}{d s}\right)^2} ds\\
\Rightarrow t(s)=&\int_{-1}^s\sqrt{\left(\tfrac{d k_x}{d s}\right)^2+\left(\tfrac{d k_y}{d s}\right)^2+\left(\tfrac{d k_z}{d s}\right)^2}ds\equiv F(s)\\
\Rightarrow s(t)=& F^{-1}(t),
\end{aligned}\label{eqt2srands2s}
\end{equation}
which cannot be computed analytically. However, the computation of $s(t)$ can be easily done numerically. Indeed, $t(s)$ is always monotonous since it is the integral of a positive function. Thus, its inverse is simply the symmetric of $F(t)$ with respect to the line of equation $t=s$. The second derivative of $\vec{k}(s)$ is given by:
\begin{equation}
\small
\begin{aligned}
\frac{d^2 k_x}{d s^2}&=\tfrac{\sin\theta_T}{2}\left(P_{n}^{(0,1)}(s)+P_{n+1}^{(0,1)}(s)\right)\\
& +\tfrac{(1+s)\sin\theta_T}{4}\left((n+2)P_{n-1}^{(1,2)}(s)+(n+3)P_{n}^{(1,2)}(s)\right),\\
\frac{d^2 k_y}{d s^2}&=\tfrac{n+1}{2}\left(P_{n}^{(0,1)}(s)-P_{n+1}^{(0,1)}(s)\right)\\
& +\tfrac{(1+s)(n+1)}{4}\left((n+2)P_{n-1}^{(1,2)}(s)-(n+3)P_{n}^{(1,2)}(s)\right)\\
\frac{d^2 k_z}{d s^2}&=-2\sin^2\left(\tfrac{\theta_T}{2}\right).
\end{aligned}
\end{equation}
and the third derivative is:
\begin{equation}
\small
\begin{aligned}
\frac{d^3 k_x}{d s^3}&=\tfrac{\sin\theta_T}{2}\left((n+2)P_{n-1}^{(1,2)}(s)+(n+3)P_{n}^{(1,2)}(s)\right)+\\
&\tfrac{(1+s)(n+3)\sin\theta_T}{8}\left((n+2)P_{n-2}^{(2,3)}(s)+(n+3)P_{n-1}^{(2,3)}(s)\right),\\
\frac{d^3 k_y}{d s^3}&=\tfrac{(n+1)}{2}\left((n+2)P_{n-1}^{(1,2)}(s)-(n+3)P_{n}^{(1,2)}(s)\right)+\\
& \tfrac{(1+s)(n+3)(n+1)}{8}\left((n+2)P_{n-2}^{(2,3)}(s)-(n+3)P_{n-1}^{(2,3)}(s)\right),\\
\frac{d^3 k_z}{d s^3}&=0.
\end{aligned}
\end{equation}
For $n=1$, the terms $P_{n-2}^{(2,3)}$ are zero. The pulse is then given by Eq.~\eqref{eqFieldFromk2s}, which cannot be simplified. Unfortunately, we do not find any function $\vec{k}(s)$ that allows to derive a simple analytic pulse as for the inversion process. Moreover, finding polynomials that lead to some pulses of low amplitude seems to be difficult in this context. Figure~\ref{Fig5} displays the pulses that cancel the offset inhomogeneities up to the third order for a $\theta_T=90^\circ$ excitation transfer. As can be seen, the amplitude of the pulse becomes very large when the robustness increases, as compared to Fig.~\ref{fig4}.
However, note that these pulses are of finite amplitude, i.e. they are not Dirac functions.
\section{Numerical Analysis}\label{numsec}
\subsection{Comparison with GRAPE}
\begin{figure}[h!]
\centering
\includegraphics[scale=0.7]{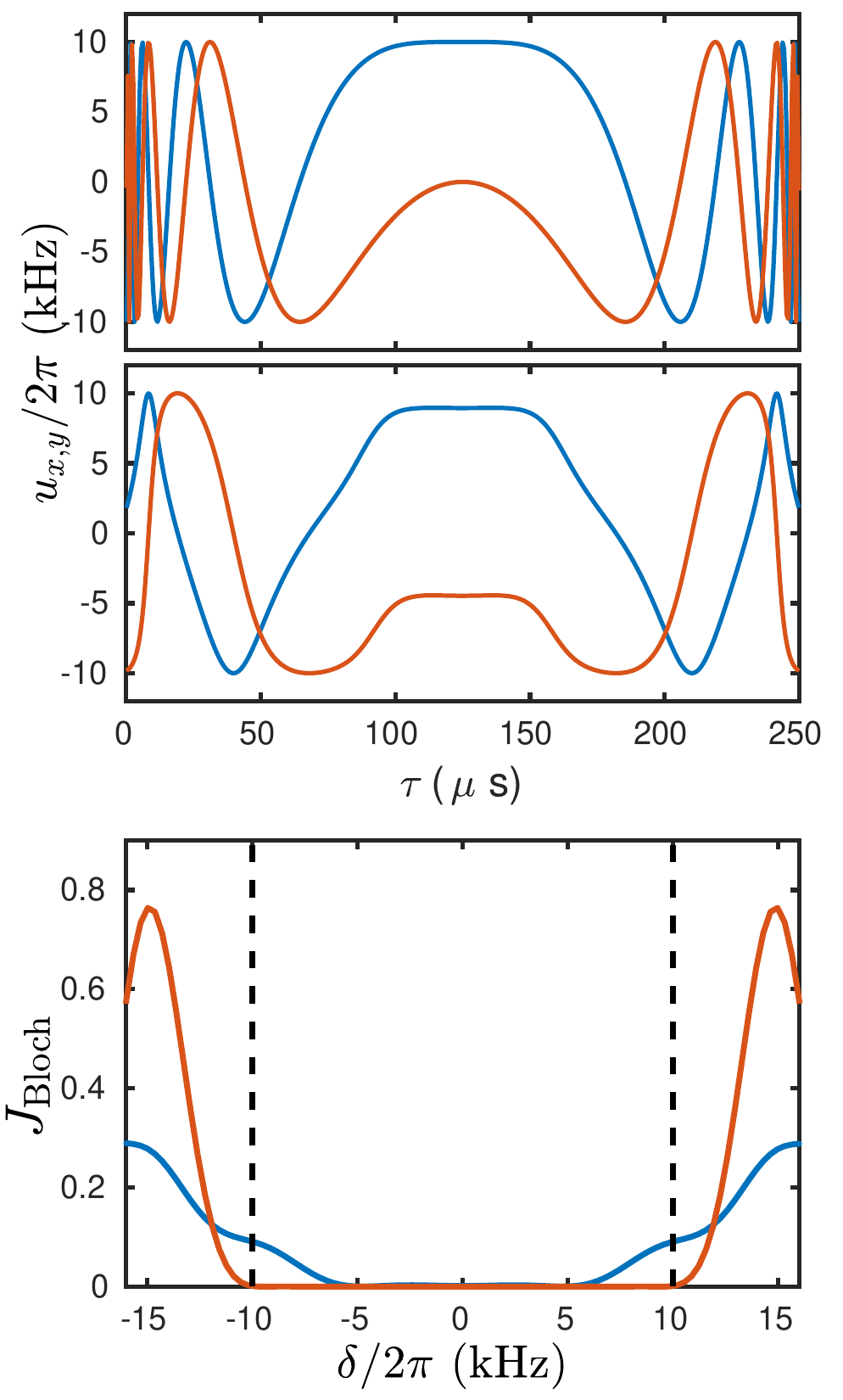}
\caption{\emph{Upper panel:} Plot of the $x$- component (blue line) and of the $y$- one (red line) of the amplitude fixed pulse given by Eq.~\eqref{PulseExp}. \emph{Middle panel:} Plot of the $x$ and $y$- components of the GRAPE pulse. \emph{Lower panel:} Cost profile associated to the amplitude-fixed pulse (blue line) and to the GRAPE pulse (red line). The dotted black lines represent the limits of the GRAPE optimization range $(\pm\omega_{\max}/2\pi)$.
 \label{fig6}}
\end{figure}

This section compares the preceding results with a robust inversion pulse optimized with the GRAPE algorithm~\cite{Khaneja05}. GRAPE is a gradient-based optimization algorithm which uses piecewise constant pulses.
We consider time steps of $\Delta\tau=0.5\,\mu$s
and amplitude-constant pulses of amplitude $\nu_{\max}=10$ kHz as in Ref.~\cite{kobzar04,kobzar12}, corresponding to:
\[\omega_{\max}=2\pi\cdot 10^4\text{ rad/s}.\]
We make a comparison with the amplitude-fixed pulse presented in Sec.~\ref{secApplic} (Eq.~\eqref{PulseSquare}), which is properly scaled by applying the formula:
\begin{equation}
\begin{aligned}
& s(\tau)=\omega_{\max}\,\tau/\sqrt{1+\nu^2},\\
& \Omega_{p}(\tau)=\omega_{\max},\\
&\phi(\tau)=\nu\ln[\sin[s(\tau)]],\\
& T_p=\pi\sqrt{1+\nu^2}/\omega_{\max}\\
\end{aligned}\label{PulseExp}
\end{equation}
where $\tau$ is the time in seconds, $T_p$ is the pulse duration and $\Omega_p$ is the pulse amplitude in rad/s. We consider a pulse of duration $T_p=250\,\mu$s by choosing $\nu=4.8990$. The numerical pulse is of the same duration and is optimized over an offset range $\delta\in[-\omega_{\max},\omega_{\max}]$. The algorithm is initialized using Eq.~\eqref{PulseExp} and is aimed to minimize the average of the cost function (Eq.~\eqref{eqCostBloch}) over the aforementioned range of offsets, i.e.:
\[\mathcal{J}=\int_{-\omega_{\max}}^{\omega_{\max}}(1+M_z(\delta,T))d\delta.\]
The resulting pulse and the cost profile are shown in Fig.~\ref{fig6}.
As expected, the numerical pulse is more efficient over the offset optimization range. However, the cost increases faster outside of this box. This feature can be explained by the fact that the amplitude-fixed pulse cannot be optimized over a certain offset range since the only degree of freedom $\nu$ is used to set the pulse duration. In contrast, GRAPE uses all the available pulse energy to improve the cost within the optimization range.
\subsection{Comparison with AHT}
\begin{figure}[h!]
\centering
\includegraphics[scale=0.75]{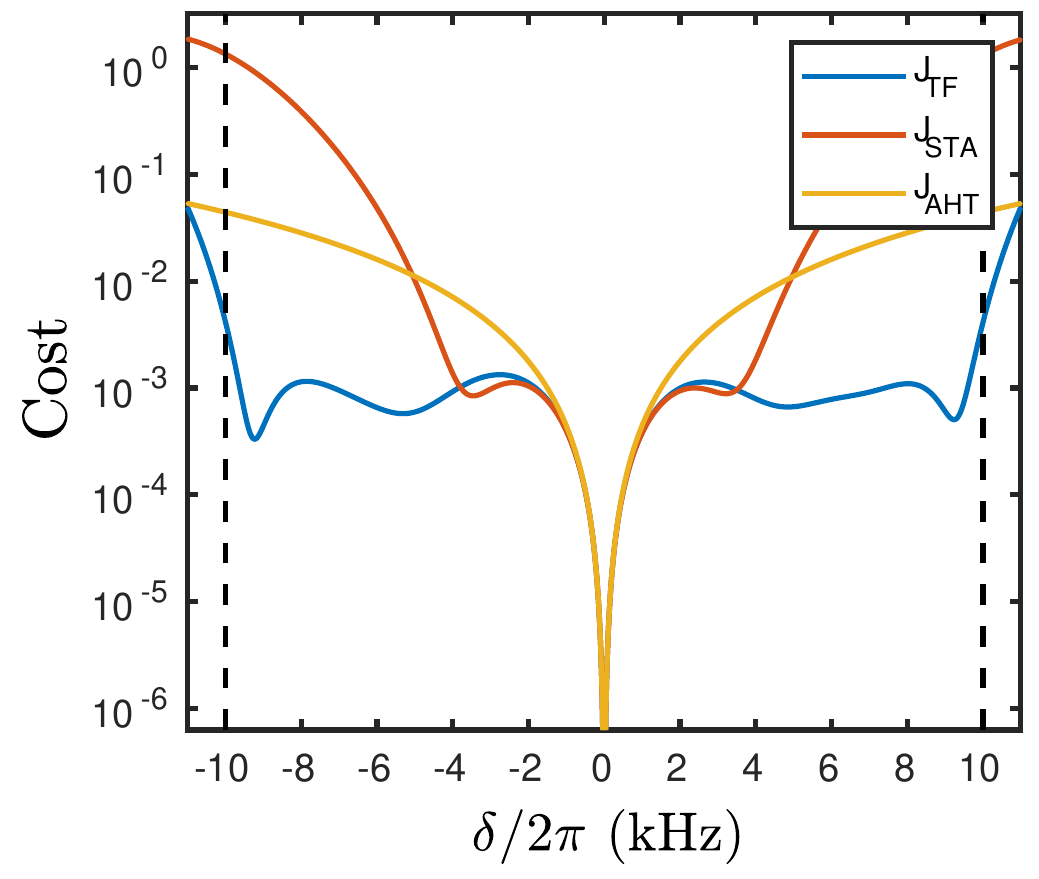}
\caption{Logarithmic view of the cost profile resulting from the GRAPE pulse, obtained using a numerical propagation and Eq.~\eqref{eqCost} (blue), using the STA approximation and Eq.~\eqref{eqCostSTA} (red) and using the AHT theory (yellow). The dotted black lines represent the limits of the optimization range of the GRAPE pulse.\label{fig7}}
\end{figure}

A very interesting question would be to know if our method could be used to derive such optimal pulses. If so, one should have to apply the Pontryagin Maximum Principle~\cite{Pontryagin} within this framework, which involves more complexity. However this issue can be partially answered by verifying the validity of our approximation for an optimal pulse by inverting the general procedure, i.e. by (i) computing $\vec{v}(t)$ numerically from the optimal pulse, (ii) computing $\ell(\delta,T)$ from Eq.~\eqref{eqSTAToggling}, (iii) computing the cost profile under STA from Eq.~\eqref{eqCostSTA} and (iv) comparing it to the cost profile measured in TF~\eqref{eqCost} determined numerically. The approximation is valid as long as the exact cost profile is well approximated. Figure~\ref{fig7} displays the cost profile of the GRAPE pulse with and without the Small Tip Angle approximation.
As can be seen in Fig.~\ref{fig6}, STA is relevant over a range $\delta/2\pi\in[-4$kHz$,4$kHz$]$ and becomes completely wrong outside this range. It does not cover the whole optimization offset range in this case. This feature however depends on the dynamics and on the pulse. A complete study would be necessary to figure out in which conditions our approximation holds. A possible approach could be to use the method of Ref.~\cite{Li17} in TF.

As a comparison, we compute the cost profile by using AHT in TF up to the first order of the Magnus expansion. The second order would have to be compared to a second order STA which would involve much more complexity. The cost profile obtained with AHT can be done by computing $\vec{L}(\delta,T)$ from $\vec{v}(t)$ through the formula:
\begin{equation}
\vec{L}(\delta,T)=U(\delta,T)\vec{L}(\delta,0)
\end{equation}
with $\vec{L}(\delta,0)=(0,0,1)^{\intercal}$ and:
\begin{equation}
U(\delta,T)=\exp\left[\delta \int_0^T \tilde{H}_1(t)dt\right],
\end{equation}
where $\tilde{H}_1(t)$ is the $3\times 3$ matrix given by Eq.~\eqref{eqHI}. The cost profile is then given by $J_{\textsc{aht}}(\delta)=1-L_z(\delta,T)$.
We can see on Fig.~\ref{fig7} that STA is valid over a larger range of offset than AHT, the latter being relevant for $\delta/2\pi\in[-2$kHz$,2$kHz$]$. STA is thus a better tool here, but is still not enough to study the properties of the pulse over the whole optimization range. In fact, our results suggest that STA is more relevant for long duration robust pulses, while AHT theory is a better approximation for short pulses.
\section{Conclusion}\label{conclusion}
We apply in this study the Small Tip-Angle Approximation in the Toggling Frame for the design of robust state to state transfers against offset inhomogeneities. Despite the apparent complexity of the method, a series of analytic and experimentally relevant pulses have been derived. Even if the final state of each system cannot be generally found analytically due to the complexity of the dynamics, the application of STA allows us to express the cost profile through a simple integral that can be computed explicitly in some cases. This approximation is particularly relevant for local robustness (small offset range) and for long duration pulses with a good performance over a large offset range. Although only a few pulses are explicited in this paper, the method can be used to derive an infinite number of analytic pulses with different properties that could be adapted to specific experimental constraints. Other state to state transfers than the inversion process bring up additional difficulties for calculating analytic pulses with a reasonable amplitude.

While STA is well known for state to state transfer in quantum control and in NMR, it could also be used here for unitary transformations. Indeed, it allows to derive the flip angle and the azimuthal angle as a function of time, that is two of the three required angles for deriving the propagator of the transformation. The third angle can be expressed as a function of the two other angles at least implicitly through an integral. Thus, the application to unitary transformation would involve more complexity, but could also be a better approximation than AHT theory. Another extension of this approach is to consider $B_1$-field inhomogeneities. In this case, the function $\vec{v}(t)$ is replaced by other functions with different properties, which stems from the fact that the interaction Hamiltonian changes. A more complex study would be to apply this method to coupled spins system, similarly to Ref.~\cite{Buterakos21} for state to state control problems.

\noindent\textbf{Acknowledgments.}\\
This research project has received funding from the European Union's Horizon 2020 research and innovation programme under the Marie-Sklodowska-Curie Grant Agreement No. 765267 (QUSCO). S.J.G. and L.V.D. acknowledge support from the Deutsche Forschungsgemeinschaft (DFG, German Research
Foundation) under Germany’s Excellence Strategy, Grant No.EXC-2111–390814868.

\appendix
\section{Derivation of equations~\eqref{eqFieldFromk2s}\label{AppEq23}}
These equations are obtained from Eq.~\eqref{eqFieldFromv}. Using the fact that $\vec{v}=\frac{d\vec{k}}{dt}=\frac{d\vec{k}}{ds}\frac{ds}{dt}$, we have:
\begin{equation}
\frac{d\vec{v}}{dt}=\left.\left.\frac{d}{dt}\right(\frac{d\vec{k}}{ds}\frac{ds}{dt}\right)=\left.\left.\frac{d}{ds}\right(\frac{d\vec{k}}{ds}\frac{ds}{dt}\right)\frac{ds}{dt}.
\end{equation}
Since $\frac{ds}{dt}=\Vert\frac{d\vec{k}}{ds}\Vert^{-1}$ (Eq.~\eqref{eqdssurdt}), we arrive at:
\begin{align}
\frac{d\vec{v}}{dt}&=\left.\left.\frac{d}{ds}\right(\frac{d\vec{k}}{ds}\Big\Vert\frac{d\vec{k}}{ds}\Big\Vert^{-1}\right)\Big\Vert\frac{d\vec{k}}{ds}\Big\Vert^{-1}\nonumber\\
& =\left(\Big\Vert\frac{d\vec{k}}{ds}\Big\Vert^{-1}\frac{d^2\vec{k}}{ds^2}-\left(\frac{d}{ds}\Big\Vert\frac{d\vec{k}}{ds}\Big\Vert\right)\Big\Vert\frac{d\vec{k}}{ds}\Big\Vert^{-2}\frac{d\vec{k}}{ds}\right)\Big\Vert\frac{d\vec{k}}{ds}\Big\Vert^{-1}
\end{align}
Using $\frac{d}{ds}\left\Vert\frac{d\vec{k}}{ds}\right\Vert=\left(\frac{d\vec{k}}{ds}\cdot\frac{d^2\vec{k}}{ds}\right)\left\Vert\frac{d\vec{k}}{ds}\right\Vert^{-1}$, we get:
\begin{align}
&\frac{d\vec{v}}{dt}=\left( \Big\Vert\frac{d\vec{k}}{ds}\Big\Vert^{-1}\frac{d^2\vec{k}}{ds^2} - \left(\frac{d\vec{k}}{ds}\cdot \frac{d^2\vec{k}}{ds} \right)\Big\Vert\frac{d\vec{k}}{ds}\Big\Vert^{-3}\frac{d\vec{k}}{ds}\right)\Big\Vert\frac{d\vec{k}}{ds}\Big\Vert^{-1}\nonumber\\
&=\left( \Big\Vert\frac{d\vec{k}}{ds}\Big\Vert^{2}\frac{d^2\vec{k}}{ds^2} - \left(\frac{d\vec{k}}{ds}\cdot \frac{d^2\vec{k}}{ds} \right)\frac{d\vec{k}}{ds}\right)\Big\Vert\frac{d\vec{k}}{ds}\Big\Vert^{-4}\nonumber\\
&=\left( \left(\frac{d\vec{k}}{ds}\cdot\frac{d\vec{k}}{ds}\right)\frac{d^2\vec{k}}{ds^2} - \left(\frac{d\vec{k}}{ds}\cdot \frac{d^2\vec{k}}{ds} \right)\frac{d\vec{k}}{ds}\right)\Big\Vert\frac{d\vec{k}}{ds}\Big\Vert^{-4}.
\end{align}
From the vector triple product, the solution becomes:
\begin{equation}
\frac{d\vec{v}}{dt}=\left[\frac{d\vec{k}}{ds}\times\left(\frac{d^2\vec{k}}{ds}\times\frac{d\vec{k}}{ds}\right)\right]\Big\Vert\frac{d\vec{k}}{ds}\Big\Vert^{-4}.
\end{equation}
Since $\frac{d\vec{k}}{ds}$ is orthogonal to  $\left(\frac{d^2\vec{k}}{ds}\times\frac{d\vec{k}}{ds}\right)$, we have $\left\Vert\frac{d\vec{k}}{ds}\times\left(\frac{d^2\vec{k}}{ds}\times\frac{d\vec{k}}{ds}\right)\right\Vert=\left\Vert\frac{d\vec{k}}{ds}\right\Vert\left\Vert\frac{d^2\vec{k}}{ds}\times\frac{d\vec{k}}{ds}\right\Vert$ and we obtain:
\begin{equation}
\Omega(s(t))=\left\Vert\frac{d\vec{v}}{dt}\right\Vert=\left\Vert\frac{d^2\vec{k}}{ds^2}\times\frac{d\vec{k}}{ds}\right\Vert\left\Vert\frac{d\vec{k}}{ds}\right\Vert^{-3},
\end{equation}
according to the first equation of~\eqref{eqFieldFromk2s}.

The derivation of the pulse's phase requires to compute the second derivative $\ddot{\vec{v}}$. We have:
\begin{align}
&\frac{d^2\vec{v}}{dt^2}=\frac{d}{ds}\Big(\frac{d\vec{v}}{dt}\Big)\frac{ds}{dt}\\
&=\frac{d}{ds}\left[\frac{d\vec{k}}{ds}\times\left(\frac{d^2\vec{k}}{ds^2}\times\frac{d\vec{k}}{ds}\right)\right]\Big\Vert\frac{d\vec{k}}{ds}\Big\Vert^{-5}\\
&=\left[\frac{d^2\vec{k}}{ds^2}\times\left(\frac{d^2\vec{k}}{ds^2}\times\frac{d\vec{k}}{ds}\right)+\frac{d\vec{k}}{ds}\times\left(\frac{d^3\vec{k}}{ds^3}\times\frac{d\vec{k}}{ds}\right)\right]\Big\Vert\frac{d\vec{k}}{ds}\Big\Vert^{-5}.
\end{align}
On the other hand, we have:
\begin{align}
&\vec{v}\times\dot{\vec{v}}=\frac{d\vec{k}}{ds}\times\left[\frac{d\vec{k}}{ds}\times\left(\frac{d^2\vec{k}}{ds^2}\times\frac{d\vec{k}}{ds}\right)\right]\left\Vert\frac{d\vec{k}}{ds}\right\Vert^{-5}\nonumber\\
&=\Bigg[\underbrace{\left[\tfrac{d\vec{k}}{ds}\cdot\left(\tfrac{d^2\vec{k}}{ds^2}\times\tfrac{d\vec{k}}{ds}\right)\right]}_{=0}\tfrac{d\vec{k}}{ds}-\left\Vert\tfrac{d\vec{k}}{ds}\right\Vert^2\left(\tfrac{d^2\vec{k}}{ds^2}\times\tfrac{d\vec{k}}{ds}\right)\Bigg]\left\Vert\tfrac{d\vec{k}}{ds}\right\Vert^{-5}\nonumber\\
&=\left(\frac{d\vec{k}}{ds}\times\frac{d^2\vec{k}}{ds^2}\right)\left\Vert\frac{d\vec{k}}{ds}\right\Vert^{-3}.
\end{align}
Thus:
\begin{equation}
\left(\vec{v}\times\dot\vec{v}\right)\cdot\ddot{\vec{v}}=\left(\tfrac{d\vec{k}}{ds}\times\tfrac{d^2\vec{k}}{ds^2}\right)\cdot\left[\tfrac{d\vec{k}}{ds}\times\left(\tfrac{d^3\vec{k}}{ds^3}\times\tfrac{d\vec{k}}{ds}\right)\right]\Big\Vert\tfrac{d\vec{k}}{ds}\Big\Vert^{-8}.
\end{equation}
From the vector triple product, we get:
\begin{align}
\left(\vec{v}\times\dot\vec{v}\right)\cdot\ddot{\vec{v}}&=\tfrac{d\vec{k}}{ds}\cdot\left[\left(\tfrac{d^3\vec{k}}{ds^3}\times\tfrac{d\vec{k}}{ds}\right)\times\left(\tfrac{d^2\vec{k}}{ds^2}\times\tfrac{d\vec{k}}{ds}\right)\right]\Big\Vert\tfrac{d\vec{k}}{ds}\Big\Vert^{-8}\nonumber\\
&=\left[\left(\tfrac{d\vec{k}}{ds}\times\tfrac{d^2\vec{k}}{ds^2}\right)\cdot\tfrac{d^3\vec{k}}{ds^3}\right]\left(\tfrac{d\vec{k}}{ds}\cdot\tfrac{d\vec{k}}{ds}\right)\Big\Vert\tfrac{d\vec{k}}{ds}\Big\Vert^{-8}\nonumber\\
&=\left[\left(\tfrac{d\vec{k}}{ds}\times\tfrac{d^2\vec{k}}{ds^2}\right)\cdot\tfrac{d^3\vec{k}}{ds^3}\right]\Big\Vert\tfrac{d\vec{k}}{ds}\Big\Vert^{-6}.
\end{align}
Thus:
\begin{equation}
\frac{\left(\vec{v}\times\dot\vec{v}\right)\cdot\ddot{\vec{v}}}{\Omega^2}=
\frac{\left(\tfrac{d\vec{k}}{ds}\times\tfrac{d^2\vec{k}}{ds^2}\right)\cdot\tfrac{d^3\vec{k}}{ds^3}}{\left\Vert\frac{d^2\vec{k}}{ds^2}\times\frac{d\vec{k}}{ds}\right\Vert^2}.
\end{equation}
Finally, we obtain:
\begin{align}
\phi(s(t))&=\int_0^t \frac{\left(\tfrac{d\vec{k}}{ds}\times\tfrac{d^2\vec{k}}{ds^2}\right)\cdot\tfrac{d^3\vec{k}}{ds^3}}{\left\Vert\frac{d^2\vec{k}}{ds^2}\times\frac{d\vec{k}}{ds}\right\Vert^2}dt'\\
&=\int_{s_0}^s\frac{\left(\tfrac{d\vec{k}}{ds'}\times\tfrac{d^2\vec{k}}{ds'^2}\right)\cdot\tfrac{d^3\vec{k}}{ds'^3}}{\left\Vert\frac{d^2\vec{k}}{ds'^2}\times\frac{d\vec{k}}{ds'}\right\Vert^2}\frac{dt'}{ds'}ds'\\
&=\int_{s_0}^s\frac{\left(\tfrac{d\vec{k}}{ds'}\times\tfrac{d^2\vec{k}}{ds'^2}\right)\cdot\tfrac{d^3\vec{k}}{ds'^3}}{\left\Vert\frac{d^2\vec{k}}{ds'^2}\times\frac{d\vec{k}}{ds'}\right\Vert^2}\left\Vert\frac{d\vec{k}}{ds'}\right\Vert ds',
\end{align}
according to the second equation of~\eqref{eqFieldFromk2s}.
\section{Generalized elliptic functions\label{AppElliptic}}
We recall in this paragraph standard results about elliptic functions.
\paragraph*{Jacobi elliptic functions.}
The standard elliptic functions can be defined through the Jacobi Amplitude $\am(u|m)$, where $u$ is the argument and $m$ the modulus such that $m\in[0,1]$~\cite{Abramowitz}. This function is defined as the inverse of an incomplete elliptic integral of the first kind given by:
\begin{equation}
\F(u|m)=\int_0^u\frac{d\phi}{\sqrt{1-m\sin^2\phi}}.
\end{equation}
The Jacobi amplitude $\am$ is thus related to $\F$ and $u$ via $\F(\am(u|m)|m)=u$. The complete elliptic integral $\K(m)$ can then be expressed as $\K(m)=\F(\pi/2|m)$.

\paragraph*{Generalized Jacobi Elliptic functions.}
A possible generalization of these functions can be constructed through the \emph{generalized incomplete elliptic integral of the first kind} $\F_n(u|m_1,\cdots,m_n)$ where the $m_i$'s are such that $m_i\in[0,1],~\forall i$. This integral is defined as:
\begin{equation}
\small
\begin{split}
&\F_n(u|m_1,\cdots,m_n)=\\
&\int_0^u\frac{d\phi}{\sqrt{(1-m_1\sin^2\phi) (1-m_2\sin^2\phi)\cdots(1-m_n\sin^2\phi)}}.
\end{split}
\end{equation}
The \emph{generalized Jacobi amplitude} $\am_n$ is the inverse of this integral, i.e. it is such that $\F_n\big(\am_n(u|m_1,\cdots,m_n)\big|m_1,\cdots,m_n\big)=u$. The complete version of the integral is given by $\K_n(m_1,\cdots,m_n)=\F_n(\pi/2|m_1,\cdots,m_n)$.
\section{Derivation of Eq.~\eqref{eqChebyshevPulse}~\label{AppPolynom}}
The derivation of $s(t)$ is made by using $\frac{ds}{dt}=\left\Vert \tfrac{d\vec{k}(s)}{d s}\right\Vert^{-1}$. The property of the Chebyshev polynomials $(1-s^2)U_{2n}^2+T_{2n+1}^2=1$ involves:
\begin{equation}
\begin{aligned}
\frac{ds}{dt}&=\tfrac{1}{\sqrt{\left(\tfrac{d k_x}{d s}\right)^2+\left(\tfrac{d k_y}{d s}\right)^2+\left(\tfrac{d k_z}{d s}\right)^2}}\\
&=\frac{1}{1+s^2}.
\end{aligned}
\end{equation}
Thus, we have:
\begin{equation}
t(s)=s+\tfrac{s^3}{3}+2,
\end{equation}
where the term $2$ in the right hand side of the equation ensures that $t(s_0)=t(-1)=0$. Inverting this relation, we get that $s(t)$ is given by Eq.~\eqref{eqChebyshevPulse}.

For the computation of the pulse,
we start from Eq.~\eqref{eqdkCheb} that we write here for convenience as:
\begin{equation}
\begin{aligned}
&\tfrac{d k_x}{d s}=(1-s^2)\sqrt{1-s^2}U_{2n}(s),\\
&\tfrac{d k_y}{d s}=(1-s^2)T_{2n+1}(s),\\
&\tfrac{d k_z}{d s}=-2s.
\end{aligned}
\end{equation}
Differentiating these equations with respect to $s$ leads to:
\begin{equation}
\begin{aligned}
& \tfrac{d^2k_x}{d s^2}=-\sqrt{1-s^2}\left[(2n+1)T_{2n+1}(s)+2s U_{2n}(s)\right]\\
& \tfrac{d^2k_y}{d s^2}=-2sT_{2n+1}(s)+(2n+1)(1-s^2)U_{2n}(s)\\
& \tfrac{d^2k_z}{d s^2}=-2,
\end{aligned}
\end{equation}
and differentiating one more time to:
\begin{equation}
\begin{aligned}
& \tfrac{d^3k_x}{d s^3}=\tfrac{3(2n+1)}{\sqrt{1-s^2}}sT_{2n+1}(s)-\left[(2n+1)^2+2\right]U_{2n}(s)\\
& \tfrac{d^3k_y}{d s^3}=-\left[(2n+1)^2+2\right]T_{2n+1}(s)-3(2n+1)sU_{2n}(s)\\
& \tfrac{d^3k_z}{d s^3}=0.
\end{aligned}
\end{equation}
The pulse is given by Eq.~\eqref{eqFieldFromk2s}. The cross product between the first and second derivatives of $\vec{k}$ can be expressed as:
\begin{equation}
\small
\begin{split}
&\frac{d\vec{k}}{d s}\times \frac{d^2\vec{k}}{d^2 s}=\\
&\begin{pmatrix}
-2(1+s^2)T_{2n+1}(s)+2s(2n+1)(1-s^2)U_{2n}(s)\\
\sqrt{1-s^2}\left[2s(2n+1)T_{2n+1}(s)+2(1+s^2)U_{2n}(s)\right]\\
(2n+1)(1-s^2)^{3/2}
\end{pmatrix}.
\end{split}
\end{equation}
We can show that:
\begin{equation}
\left\Vert\tfrac{d\vec{k}}{d s}\times \tfrac{d^2\vec{k}}{d^2 s}\right\Vert=(1+s^2)\sqrt{4+(2n+1)^2(1-s^2)}.
\end{equation}
The control amplitude being $\Omega(s)=\left\Vert\frac{d\vec{k}}{d s}\times \frac{d^2\vec{k}}{d^2 s}\right\Vert \left\Vert\frac{d\vec{k}}{d s}\right\Vert^{-3}$, we obtain Eq.~\eqref{eqChebyshevPulse}.

For the phase of the pulse, we have, in a first step:
\begin{equation}
\left(\tfrac{d\vec{k}}{d s}\times \tfrac{d^2\vec{k}}{d^2 s}\right)\cdot \tfrac{d^3\vec{k}}{d s^3}=-\tfrac{2ks\left[(2n+1)^2(1-s^2)+s^2+5\right]}{\sqrt{1-s^2}}.
\end{equation}
Since $d\phi/ds=\tfrac{\left(\frac{d\vec{k}}{d s}\times\frac{d^2\vec{k}}{d s^2}\right)\cdot\frac{d^3\vec{k}}{d s^3}}{\left\Vert \frac{d\vec{k}}{d s}\times\frac{d^2\vec{k}}{d s^2}\right\Vert^2} \left\Vert \tfrac{d\vec{k}}{d s}\right\Vert$, we have:
\begin{equation}
\begin{aligned}
\tfrac{d\phi}{ds}&=-\tfrac{2ks\left[(2n+1)^2(1-s^2)+s^2+5\right]}{(1+s^2)\sqrt{1-s^2}[(2n+1)^2(1-s^2)+4]}\\
 & =-\tfrac{2(2n+1)s}{\sqrt{1-s^2}[(2n+1)^2(1-s^2)+4]}-\tfrac{2(2n+1)s}{(1+s^2)\sqrt{1-s^2}}.
\end{aligned}
\end{equation}
Making the change of variable $x=s\sqrt{\frac{1-s^2}{2}}$, we obtain:
\begin{equation}
d\phi=\frac{\sqrt{2}}{2+x^2}+\frac{\sqrt{2}}{1-\frac{x^2}{(2n+1)^2}}dx.
\end{equation}
Integrating this equation and coming back to the variable $s$, we arrive at the formula of Eq.~\eqref{eqChebyshevPulse}.

\bibliographystyle{unsrt}

\end{document}